\documentclass[fleqn,usenatbib]{mnras}

\usepackage{newtxtext,newtxmath}

\usepackage[T1]{fontenc}

\DeclareRobustCommand{\VAN}[3]{#2}
\let\VANthebibliography\thebibliography
\def\thebibliography{\DeclareRobustCommand{\VAN}[3]{##3}\VANthebibliography}

\usepackage{graphicx}	
\usepackage{amsmath}	

\newcommand{\msun}{M$_{\odot}$}

\newcommand{\kms}{km~s$^{-1}$}

\newcommand{\Ha}{H$\alpha$}

\newcommand{\HeI}{He~{\sc i}}

\newcommand{\Oneb}{[O~{\sc i}]}
\newcommand{\CI}{C~{\sc i}}

\newcommand{\CaII}{Ca~{\sc ii}}

\newcommand{\FeII}{Fe~{\sc ii}}

\newcommand{\NiII}{Ni~{\sc ii}}

\newcommand{\Nifs}{$^{56}$Ni}
\newcommand{\mej}{$M_\mathrm{ej}$}
\newcommand{\ek}{$E_\mathrm{k}$}

\newcommand{\eom}{$E_\mathrm{k}/M_\mathrm{ej}$}
\newcommand{\lam}{$\lambda$}

\newcommand{\Eh}{$E\left(B-V\right)_\mathrm{host}$}
\newcommand{\Emw}{$E\left(B-V\right)_\mathrm{MW}$}
\newcommand{\Etot}{$E\left(B-V\right)_\mathrm{tot}$}

\newcommand{\mzams}{$M_\mathrm{ZAMS}$}

\newcommand{\sn}{SN\,}
\newcommand{\Oneblam}{[O~{\sc i}] \lam\lam6300, 6364}
\newcommand{\caiif}{[\CaII]}
\newcommand{\caiiflam}{[\CaII] \lam\lam7291, 7323}
\newcommand{\fo}{\caiif/\Oneb}
\newcommand{\co}{\CaII-NIR/\Oneb}
\newcommand{\fc}{\caiif/\CaII-NIR}

\newcommand{\fwhmo}{FWHM$_{6300}$}
\newcommand{\fwhmf}{FWHM$_{7921}$}
\newcommand{\fwhmc}{FWHM$_{8662}$}
\newcommand{\eomu}{[$10^{51}$ erg/\msun]}

\title[O and Ca relationships in the nebular phase]{Oxygen and calcium nebular emission line relationships in core-collapse supernovae and Ca-rich transients}

\author[S. J . Prentice.]{
S. J. Prentice,$^{1}$
K. Maguire$^{1}$\thanks{E-mail: kate.maguire@tcd.ie}, L. Siebenaler$^{1}$, and A. Jerkstrand$^{2}$
\\
$^{1}$School of Physics, Trinity College Dublin, College Green, Dublin 2, Ireland\\
$^{2}$Department of Astronomy, Oskar Klein Centre, Stockholm University, Albanova, SE-10691 Stockholm, Sweden\\
}

\date{Accepted XXX. Received YYY; in original form ZZZ}

\pubyear{2021}

\begin{document}
\label{firstpage}
\pagerange{\pageref{firstpage}--\pageref{lastpage}}
\maketitle

\begin{abstract}
This work examines the relationships between the properties (flux ratios, full width at half-maximum velocities) of the \Oneblam, \caiiflam, and the \CaII\ near-infrared triplet, emission lines of a large sample of core-collapse supernovae (SNe) and Ca-rich transients (509 spectra of 86 transients, of which 10 transients are Ca-rich events). Line-flux ratios as a function of time were investigated with differences identified between the transient classes, in particular the Type II SNe were found to have distinct line-flux ratios compared to stripped-envelope (SE) SNe. No correlation was found between the \fo\ flux ratios of SE-SNe and their ejecta masses and kinetic energies (as measured from light curve modelling), suggesting that there may be a contribution from an additional power source in more luminous SE-SNe. We found that the mean characteristic width of the \caiif\ emission line is less than the \Oneb\ emission line for all SN types, indicating that the \caiif\ emission typically originates from deeper in the ejecta than \Oneb. This is in some tension with standard models for emission in Type II SNe. The emission line properties of Type II SNe were also compared to theoretical models and found to favour lower mass tracks (\mzams $<$ 15 \msun), with no evidence found for significant mixing of \Nifs\ into the H envelope nor Ca mixed into the O shell.
The flux ratios of some superluminous SNe were found to be similar to those of SE-SNe when scaling to account for their longer rise times was applied (although we caution the sample size is small). 

\end{abstract}





\section{Introduction}

The nebular phase of supernovae (SNe) represents the epoch ($\gtrsim$ 100 d after explosion) when the ejecta become largely optically thin, giving a view into the inner workings of the explosion. Spectra at these times are dominated by emission lines, the type and strength of which depends on SN type. Core-collapse SNe are thought to result from the explosions of massive stars with masses of $>$8 \msun\ \citep[e.g.][]{2003ApJ...591..288H}.
In normal core-collapse SNe (CC-SNe), these lines are primarily \Ha, present in Type II SNe (SNe II) but not in stripped-envelope (SE) Type Ibc SNe, as well as \Oneblam, \caiiflam, and \CaII\ NIR   \lam \lam8498, 8542, 8662 \citep{1997ARA&A..35..309F} lines. Type IIb SNe are those that look like a SN II (H-rich) at early times but evolve to look more like a Type Ib (He-rich) within a few weeks of explosion \citep[e.g.][]{1988AJ.....96.1941F,1993Natur.364..507N}. Broad-lined Type Ic (Ic-BL) are a specific class of SE-SNe that show lines with very high expansion velocities (15,000 -- 30,000 \kms\ at peak)  compared to normal Type Ic  that have mean velocities of $\sim$10,000 \kms at peak \citep{2014AJ....147...99M}.  Long-duration gamma ray bursts have been associated with some (but not all) of these Ic-BL events \citep{2006ARA&A..44..507W}. Superluminous SNe (SLSNe) are many times ($>$10) more luminous than normal CC-SNe and come in two classes, those that show signatures of H-interaction at late times (SLSN-II) and those that do not  \citep[SLSN-I;][]{2011Natur.474..487Q,2019ARA&A..57..305G}. `Ca-rich' transients are named for their strong \caiif\ and \CaII\ NIR emission lines at late times, although this does not necessarily directly link to the amount of Ca present in their ejecta \citep[e.g.][]{2010Natur.465..322P} and their origin remains unclear with both thermonuclear and CC progenitors proposed \citep[e.g.][]{2017ApJ...846...50M, 2020ApJ...905...58D}. In thermonuclear SNe (SNe Ia), the dominant emission in the nebular phase is from Fe-group elements but in a few rare cases, an \Oneb\ line has been identified \citep{2013ApJ...775L..43T,2016MNRAS.459.4428K}. Ca-emission lines are also rare in SNe Ia with only a few detections of the \caiif\ and \CaII\ NIR triplet lines in some peculiar events \citep[e.g.][]{2007A&A...475..585A,2013ApJ...775L..43T, 2020ApJ...900L..27S}. 

The spectra of interacting SNe display strong narrow emission lines; \Ha\ for Type IIn and \HeI\ and Fe lines for Type Ibn (SNe Ibn). These narrow lines are due to interaction between the SN ejecta and nearby circumstellar material (CSM). Interacting SNe may also show strong emission from Ca features and in rare cases from weak \Oneb. The well-studied SN Ibn, SN\,2006jc \citep{2007ApJ...657L.105F,2007Natur.447..829P}, for example, displays strong \HeI\ emission throughout its evolution but three months after discovery, the \caiif\ line appears and very soon after a weak \Oneb\ line can be identified.

The prominence of emission lines in the nebular phase
allow the ejecta distribution and signs of asymmetry to be inferred through measured properties such as the intensity, flux/luminosity, and full width at half maximum  \citep[FWHM; e.g.][]{2009MNRAS.397..677T,2012MNRAS.420.3451M,2018MNRAS.476.2905M,2022arXiv220111467F}. A full investigation of the properties of the inner ejecta of SNe requires modelling with radiative-transfer codes owing to the non-local thermodynamic equilibrium (non-LTE) nature of the line-forming processes during the nebular phase.
Such methods have been applied in the literature to SNe II \citep[e.g.,][]{Dessart2011,Jerkstrand2012,Jerkstrand2014}, SNe Ia \citep[e.g.,][]{Botyanszki2018,2020MNRAS.494.2809M,2020MNRAS.492.2029S}, SNe Ibc \citep[e.g.,][]{Mazzali2009,Mazzali2010,Jerkstrand2015,2020MNRAS.497.3542P}, and SLSNe \citep[e.g.][]{Dessart2013,Jerkstrand2020}.
These methods help to provide a physical interpretation to the results of analytical methods. 
For example, the luminosity \citep{Jerkstrand2014,Jerkstrand2015} and width \citep{2010MNRAS.408..827D} of the \Oneb\ line has been used to estimate the O masses of the progenitor stars in CC-SNe.

Another approach for quantifying SN properties, which is commonly used for Ca-rich transients \citep[e.g.][]{Valenti2014,2017ApJ...846...50M}, is to measure the flux ratio of the \caiif\ and \Oneb\ emission features. The origin of Ca-rich transients is unknown and it is likely that several sub-populations exist \citep{2020ApJ...905...58D}. The late-phase spectra of Ca-rich transients are akin to Type Ibc with very weak \Oneb\ emission, and their \caiif-to-\Oneb\ ratio has been used to characterise their nebular spectra in relation to CC-SNe. 
However, in some cases their preference for occurring in remote locations in galaxy clusters is more suggestive of white-dwarf progenitors \citep{2018ApJ...858...50F} rather than massive star explosions.

Emission from \caiif, \CaII\ NIR, and \Oneb\ arise from different regions of the ejecta depending on the SN type.
In H-rich SNe (Type II), the Ca emission lines are predicted to arise primarily from primordial Ca in the inner parts of the H envelope, with emission from synthesised Ca being insufficient to match observations \citep{Li1993,KF98,Jerkstrand2012}.
Forbidden \CaII\ lines form in the inner denser regions, while the \CaII\ NIR triplet is proposed to have contributions from throughout the envelope \citep{Dessart2011}. 
Note, however, the models by \citet{2021A&A...652A..64D} in which Ca ionizes to Ca III in the H envelope.
In SE-SNe (Type Ibc), the lack of a H envelope means that the source of \caiif\ and \CaII\ lines is through explosive O burning, or incomplete explosive Si burning and \Nifs\ burning, respectively. 
In the SN IIb (SE-SNe with small amounts of H) nebular models of \citet{Jerkstrand2015}, the \CaII\ NIR lines are formed by fluorescence from H\&K UV pumping in trace Ca in the $^{56}$Ni-zone, showing how complex the line formation can be. Finally, using measurements of Ca synthesised in explosive burning for estimating the zero-age main sequence mass (\mzams) of the progenitor star is difficult because the amount of Ca made depends on the explosion energy as function of progenitor mass, a relationship that we currently have little knowledge about \citep{Jerkstrand2017}.

The aim of this work is to analyse a large sample of nebular-phase spectra using empirical methods to characterise the \caiif, \CaII\ NIR, and \Oneb\ emission lines of a range of SN types and connect them to their physical properties and underlying explosions. 
Nebular spectral models of SNe II and SNe IIb from \citet{Jerkstrand2014,Jerkstrand2015} are included in the analysis for comparison.
In Section~\ref{sec:sample}, the sample is presented along with a description of how the spectra are prepared for this analysis, and in Section~\ref{sec:method} the line-fitting method is described.
The results are presented in Section~\ref{sec:results} and discussed in Section~\ref{sec:discussion}, along with a comparison of the observed spectra to models.
Finally, the conclusions are presented in Section~\ref{sec:conclusions}.

\section{Sample collection and pre-analysis}\label{sec:sample}

In this section, we present the details of the spectral sample used and estimates of their spectral phases, as well as an overview and description of the prominent lines in the different SN types used in this analysis.

\begin{table*}
    \centering
    \caption{The SN sample and their properties.}
    \begin{tabular}{lcccccc}
    \hline
    SN & Type$^\ddag$ & $z$ & \Emw  & \Eh$\dag$ & Epochs & References  \\
     & & & [mag] & [mag] & [days] & \\
    \hline
1987A & II  & 1e-05 & 0.01 &  -  & 113-445 &  1  \\
1990B & Ic  & 0.0075 & 0.03 &  0.5  & 83-95 &  2, 3  \\
1990E & II  & 0.0042 & 0.02 &  0.5  & 133-347 &  4, 5, 6, 7  \\
1990K & II  & 0.0052 & 0.5 &  -  & 119-203 &  8, 7  \\
1992H & II  & 0.0095 & 0.09 &  -  & 82-467 &  7, 9, 10, 6  \\
1993J & IIb  & -0.0001 & 0.07 &  0.1  & 86-500 &  11, 12, 13, 14, 15  \\
1993K & II  & 0.0094 & 0.05 &  -  & 223-307 &  7, 16  \\
1994I & Ic  & 0.0018 & 0.03 &  0.3  & 53-144 &  12, 17  \\
1996cb & IIb  & 0.0024 & 0.12 &  -  & 91-332 &  12, 3  \\
1997ef & Ic-BL  & 0.012 & 0.04 &  -  & 88-295 &  3, 12  \\
1998bw & Ic-BL  & 0.0085 & 0.05 &  -  & 120-332 &  18  \\
1999em & II  & 0.0023 & 0.1 &  0.001  & 116-371 &  19, 20, 16  \\
2001ig & IIb  & 0.0031 & 0.02 &  -  & 282-313 &  21  \\
2002ap & Ic-BL  & 0.0021 & 0.07 &  0.008  & 113-376 &  22, 23, 12  \\
2002hh & II  & 0.0001 & 1.96 &  -  & 154-389 &  20  \\
2003B & II  & 0.0036 & 0.02 &  -  & 247-247 &  16  \\
2003bg & IIb  & 0.0043 & 0.02 &  -  & 130-286 &  24  \\
2003hn & II  & 0.0059 & 0.01 &  0.18  & 86-160 &  16  \\
2004aw & Ic  & 0.016 & 0.02 &  0.35  & 54-399 &  25, 13  \\
2004dj & II  & 0.0004 & 0.1 &  0.001  & 56-399 &  7, 26, 27, 28  \\
2004et & II  & 0.0013 & 0.04 &  0.001  & 88-384 &  29, 20  \\
2004gk & Ic  & -0.0006 & 0.03 &  0.1  & 74-223 &  13, 30  \\
2004gq & Ib  & 0.0064 & 0.06 &  0.10  & 80-343 &  12, 30  \\
2004gt & Ib  & 0.0054 & 0.04 &  0.2  & 151-151 &  23  \\
2005E & Ca-rich  & 0.0089 & 0.09 &  -  & 17-50 &  31  \\
2005ay & II  & 0.0026 & 0.1 &  -  & 249 &  20  \\
2005bf & Ib  & 0.0189 & 0.04 &  -  & 201-209 &  12, 13  \\
2005cs & II  & 0.002 & 0.05 &  -  & 86-254 &  32, 20  \\
2006aj & Ic-BL  & 0.033 & 0.09 &  -  & 195-282 &  12, 30, 23  \\
2006bp & II  & 0.0035 & 0.02 &  0.2  & 329 &  33  \\
2006el & IIb  & 0.017 & 0.09 &  -  & 84 &  13  \\
2006jc & Ibn  & 0.0055 & 0.15 &  -  & 56-155 &  34, 35, 12, 36, 37, 13  \\
2007I & Ic-BL  & 0.0216 & 0.02 &  -  & 148-175 &  23  \\
2007aa & II  & 0.0048 & 0.02 &  -  & 336 &  38  \\
2007gr & Ic  & 0.0017 & 0.05 &  0.03  & 64-417 &  13  \\
2007ke & Ca-rich  & 0.0173 & 0.09 &  -  & 18-18 &  39  \\
2008D & Ib  & 0.0065 & 0.02 &  0.063  & 87-127 &  40  \\
2008ax & IIb  & 0.0019 & 0.02 &  0.28  & 93-390 &  41, 12, 42, 43  \\
2008bk & II  & 0.0007 & 0.02 &  -  & 139-429 &  16, 38  \\
2008bo & IIb  & 0.005 & 0.05 &  0.03  & 83-194 &  13, 12  \\
2009N & II  & 0.0034 & 0.01 &  0.11  & 84-356 &  44, 38  \\
PTF09dav & Ca-rich  & 0.0371 & 0.04 &  -  & 92 &  45  \\
2009ib & II  & 0.0043 & 0.09 &  0.5  & 60-179 &  46  \\
2009jf & Ib  & 0.0079 & 0.11 &  0.05  & 53-389 &  13, 47, 12  \\
2010et & Ca-rich  & 0.023 & 0.04 &  0.001  & 60-112 &  45  \\
2010lp & Ia  & 0.01 & 0.21 &  -  & 264 &  48  \\
2010md & SLSN*  & 0.0982 & 0.09 &  -  & 241-315 &  49, 50  \\
PTF11bij & Ca-rich  & 0.035 & 0.01 &  -  & 41 & 51 \\
2011bm & Ic  & 0.022 & 0.03 &  0.032  & 96-257 & 52 \\
2011dh & IIb  & 0.0016 & 0.03 &  0.05  & 60-376 & 53,54,55 \\
2011ei & IIb  & 0.0093 & 0.05 &  0.18  & 97-310 & 56 \\
2011fu & IIb  & 0.0184 & 0.06 &  0.015  & 62-274 & 57,58, 13  \\
2011hs & IIb  & 0.0057 & 0.01 &  0.16  & 158-332 & 59 \\
PTF11kmb & Ca-rich  & 0.017 & 0.09 &  -  & 62-122 & 60 \\
OGLE-2012-SN-006 & Ibn  & 0.06 & 0.06 &  -  & 66-160 & 61 \\
2012ap & Ic-BL  & 0.0122 & 0.04 &  0.4  & 209-263 &  13,62 \\
2012aw & II  & 0.0025 & 0.03 &  0.04  & 227-383 & 63, 7  \\

    \hline   
    \end{tabular}
    
    \label{tab:sample}
\end{table*}

\begin{table*}
    \centering
    \contcaption{The SN sample and their properties.}
    \begin{tabular}{lcccccc}
    \hline
    SN & Type$^\ddag$ & $z$ & \Emw  & \Eh$\dag$ & Epochs & References  \\
     & & & [mag] & [mag] & [days] & \\
    \hline
PTF12bho & Ca-rich  & 0.023 & 0.0 &  -  & 16-134 & 64 \\
PTF12dam & SLSN-I  & 0.1075 & 0.01 &  -  & 264-319 &  50  \\
2012ec & II  & 0.0046 & 0.08 &  0.001  & 145-384 &  7  \\
2012hn & Ca-rich  & 0.0076 & 0.3 &  0.001  & 148 & 65 \\
2013ab & II  & 0.0053 & 0.04 &  -  & 68-175 &  7  \\
2013ak & IIb  & 0.0037 & 0.4 &  -  & 177 & 66 \\
2013am & II  & 0.0026 & 0.09 &  0.55  & 186-410 &  7  \\
2013bb & IIb  & 0.0175 & 0.01 &  0.2  & 316 & 67 \\
iPTF13bvn & Ib  & 0.0044 & 0.02 &  0.0437  & 287 & 68 \\
2013ej & II  & 0.0021 & 0.07 &  -  & 51-412 & 69,70 \\
2013ge & Ib/c  & 0.0043 & 0.02 &  0.047  & 66-420 &  13,71 \\
2014G & II  & 0.0045 & 0.21 &  -  & 88-327 & 72 \\
2014L & Ic  & 0.008 & 0.03 &  0.5  & 54-140 &  13  \\
2015G & Ibn  & 0.0047 & 0.4 &  -  & 99-161 & 73 \\
2015ah & Ib  & 0.016 & 0.07 &  0.02  & 146 & 67 \\
2015ap & Ib  & 0.0113 & 0.03 &  -  & 101-141 & 67 \\
PS15bgt & IIb  & 0.0089 & 0.05 &  0.25  & 83-140 &  13  \\
2015bn & SLSN-I  & 0.1136 & 0.02 &  -  & 256-389 & 74,75 \\
iPTF15eqv & Ca-rich  & 0.0046 & 0.02 &  -  & 71-223 & 76 \\
2016coi & Ic-BL  & 0.0036 & 0.08 &  0.125  & 151-440 & 77 \\
2016eay & SLSN-I  & 0.1013 & 0.01 &  -  & 153-199 & 78,79 \\
2016hgs & Ca-rich  & 0.017 & 0.05 &  -  & 26-55 & 80 \\
2016iae & Ic  & 0.0036 & 0.01 &  0.65  & 89-103 & 67 \\
2016jdw & Ib  & 0.0189 & 0.01 &  -  & 101-126 & 67 \\
2017bgu & Ib  & 0.0085 & 0.02 &  0.02  & 114-183 & 67 \\
2017egm & SLSN-I  & 0.0307 & 0.0 &  -  & 350 & 81 \\
2017ein & Ic  & 0.0027 & 0.02 &  0.4  & 187-192 & 82 \\
2018gjx & Ibn  & 0.012 & 0.05 &  0.012  & 46-141 & 83 \\
2019yz & Ic  & 0.006 & 0.1 &  0.2  & 121-173 & 84 \\
    \hline  
    \multicolumn{7}{p{\textwidth}}{References: (1) \cite{1995ApJS...99..223P}, (2) \cite{2001ApJ...553..886C}, (3) \cite{2001AJ....121.1648M}, (4) \cite{1994A+A...285..147B}, (5) \cite{1993AJ....105.2236S}, (6) \cite{2000AJ....120..367G}, (7) \cite{2017MNRAS.467..369S}, (8) \cite{1995A+A...293..723C}, (9) \cite{1996AJ....111.1286C}, (10) \cite{1997ARA+A..35..309F}, (11) \cite{2000AJ....120.1499M}, (12) \cite{2014AJ....147...99M}, (13) \cite{2019MNRAS.482.1545S}, (14) \cite{1995A+AS..110..513B}, (15) \cite{Jerkstrand2015}, (16) \cite{2017ApJ...850...89G}, (17) \cite{1995ApJ...450L..11F}, (18) \cite{2001ApJ...555..900P}, (19) \cite{2002PASP..114...35L}, (20) \cite{2014MNRAS.442..844F}, (21) \cite{2009PASP..121..689S}, (22) \cite{2003PASP..115.1220F}, (23) \cite{2009MNRAS.397..677T}, (24) \cite{2009ApJ...703.1612H}, (25) \cite{2006MNRAS.371.1459T}, (26) \cite{2011ApJ...732..109M}, (27) \cite{2006Natur.440..505L}, (28) \cite{2006MNRAS.369.1780V}, (29) \cite{2006MNRAS.372.1315S}, (30) \cite{2008ApJ...687L...9M}, (31) \cite{2010Natur.465..322P}, (32) \cite{2009MNRAS.394.2266P}, (33) \cite{2007ApJ...666.1093Q}, (34) \cite{2008MNRAS.389..113P}, (35) \cite{2007Natur.447..829P}, (36) \cite{2009MNRAS.392..894A}, (37) \cite{2008ApJ...680..568S}, (38) \cite{2012MNRAS.420.3451M}, (39) \cite{2019MNRAS.482.1545S}, (40) \cite{2009ApJ...702..226M}, (41) \cite{2011MNRAS.413.2140T}, (42) \cite{2011ApJ...739...41C}, (43) \cite{2010ApJ...709.1343M}, (44) \cite{2014MNRAS.438..368T}, (45) \cite{2012ApJ...755..161K}, (46) \cite{2015MNRAS.450.3137T}, (47) \cite{2011MNRAS.416.3138V}, (48) \cite{2013ApJ...775L..43T}, (49) \cite{2016ApJ...830...13P}, (50) \cite{2018ApJ...855....2Q}, (51) \cite{Lunnan2017}, (52) \cite{Valenti2012}, (53) \cite{2015A+A...580A.142E}, (54) \cite{2014A+A...562A..17E}, (55) \cite{2013MNRAS.436.3614S}, (56) \cite{2013ApJ...767...71M}, (57) \cite{2013MNRAS.431..308K}, (58) \cite{2015MNRAS.454...95M}, (59) \cite{2014MNRAS.439.1807B}, (60) \cite{Lunnan2017}, (61) \cite{2015MNRAS.449.1941P}, (62) \cite{2015ApJ...799...51M}, (63) \cite{Jerkstrand2014}, (64) \cite{Lunnan2017}, (65) \cite{Valenti2014}, (66) \cite{2015A+A...579A..40S}, (67) \cite{Prentice2019}, (68) \cite{2015A+A...579A..95K}, (69) \cite{2016PASA...33...55C}, (70) \cite{2016MNRAS.461.2003Y}, (71) \cite{2016ApJ...821...57D}, (72) \cite{2016MNRAS.462..137T}, (73) \cite{2017MNRAS.471.4381S}, (74) \cite{Nicholl2016}, (75) \cite{Jerkstrand2017}, (76) \cite{2017ApJ...846...50M}, (77) \cite{2018MNRAS.478.4162P}, (78) \cite{2017ApJ...835L...8N}, (79) \cite{2017MNRAS.469.1246K}, (80) \cite{2018ApJ...866...72D}, (81) \cite{2019ApJ...871..102N}, (82) \cite{Teffs2021}, (83) \cite{2020MNRAS.499.1450P}, (84) \cite{2021MNRAS.508.4342P}}\\
    \multicolumn{7}{p{\textwidth}}{$\dag$ Host reddening was corrected for where possible, although it has little impact on the line ratios. If no value is given then \Eh\ was either determined to be negligible or is unknown. Please consult the cited papers for uncertainties and discussion related to these values for specific objects.}\\
    \multicolumn{7}{p{\textwidth}}{*\sn2010md is an example of a SLSN-IIb  \citep[see e.g.][]{2018ApJ...855....2Q}.  }
    \\
    \end{tabular}

\end{table*}

\subsection{Sample selection and criteria}
The nebular-phase spectra used for this sample were predominantly taken from the literature and obtained from the Weizmann Interactive Supernova Data Repository\footnote{www.wiserep.org} \citep[WISeREP;][]{2012PASP..124..668Y} and the Open Supernova Catalog\footnote{https://github.com/astrocatalogs/supernovae} \citep[OSC;][]{2017ApJ...835...64G}.
Some Liverpool Telescope \citep[LT;][]{2004SPIE.5489..679S} data of recently published objects were also included. A comprehensive list of references for each event can be found in Table~\ref{tab:sample}. The criteria for selecting spectra for inclusion in this study were the following:

\begin{itemize}
    \item {Each spectrum should fully cover at least two of the three lines of interest (\caiif, \CaII\ NIR, and \Oneb) down to the continuum on both sides of the feature (discussed further below).}
    \item{The object should display emission from \Oneblam\ at some point in its evolution. Boundary cases were assessed by fitting a line to the flux in this region and required at least one three-sigma detection of the bluer of these two lines (6300 \AA) within $\pm 500$ \kms\ of the rest wavelength. If a detection was weaker than this, another spectrum was required to verify the line. In reality, no object with \Oneb\ emission presented a line this weak. }
    \item{Each spectrum should have sufficient signal-to-noise (S/N) that emission from the lines was comparable across all available spectra for an object. If the quality of a spectrum was such that the emission feature was dominated by noise (a less than three-sigma detection of the emission feature) then it was excluded from the sample.}
\end{itemize}

After applying these selection criteria, the spectra were prepared for analysis by correcting for redshift and extinction, \Etot, using the reddening law of \citet{CCM}.  Where possible the value of \Etot\ is a combination of Milky Way and host galaxy extinction  based on the values determined in the references listed in Table \ref{tab:sample}. The redshifts were also taken from the literature.
Practically, however, the correction for reddening had little impact because the relative change in flux across 6000 -- 9000 \AA\ is small for the reddening values for the sample (Table~\ref{tab:sample}). 

The phase of each spectrum was calculated by taking the observed time of the spectrum relative to the time of $r$ or $R$ maximum light of the object and converting to the rest frame using the redshift given in Table \ref{tab:sample}. If photometry in these bands was unavailable, then the following methods in order were used to estimate the phase: $i$ or $I$ maximum light, $g$  or $V$ maximum light, $B$ maximum light, or date of discovery.  Using any band other than $r$-band or $R$-band shifts the phase of the spectrum by a few days. However, this shift is negligible since we considered observations at $\sim$100 d after maximum light when any spectroscopic evolution is slow so this does not significantly impact the results of this study. 
For the time of peak, we used the typical peak from radioactive decay where present (e.g., SE-SNe, \sn1987A) and the peak due to thermal cooling of a shocked envelope in other cases (SNe II). The time difference between SE-SNe and SNe II in this context is $\sim 10$ days, but is tens of days for \sn1987A-like events. 

The names of the SNe and their classification types, as well as references for their light curves are given in Table~\ref{tab:sample}. The classification is based on the literature classification for individual events. The SE-SN classifications were also checked for consistency with the classification analysis of \cite{Prentice2017} and \citet{2017PASP..129e4201S} and were found to be consistent.
The final sample size and breakdown of SN classes is given in Table~\ref{tab:samplesize} and consists of 509 spectra of 86 transients\footnote{We have included one peculiar, underluminous and slow-evolving SN Ia, SN 2010lp \citep{2013ApJ...775L..43T} that fulfilled our selection criteria due to the presence of \Oneb\ in its late-time spectrum. However, the physical mechanism producing this emission and driving the line ratios in thermonuclear explosions is expected to be different to the other events studied here and so we do not discuss it further.}.

\begin{table}
    \centering
    \caption{Number of SN of each type used in this analysis.}
    \begin{tabular}{lc}
    \hline
     SN type    &  No. of objects \\
     \hline
     II & 24 \\
     IIb & 14 \\
     Ib & 10 \\
     Ic & 11 \\
     Ic-BL & 7 \\
     Ibn & 4 \\
     Ca-rich & 10 \\
     SLSN$^1$ & 5 \\
     Ia$^2$ & 1 \\
    \hline
        \end{tabular}
    
    \begin{flushleft}
$^1$The SLSNe in our sample are all of the SLSN-I subclass but we note that SN 2010md is an example of the rare class of SLSN-IIb.\\
$^2$The single SN Ia is a peculiar event \citep[\sn2010lp;][]{2013ApJ...775L..43T}. Normal SNe Ia do not show the \caiif, \CaII\ NIR, and \Oneb\ emission lines investigated here. \\
  \end{flushleft}

    \label{tab:samplesize}
\end{table}

\subsection{Spectral line identification in the different SN subtypes}\label{sec:pre}

\begin{figure*}
    \centering
    \includegraphics[scale=0.65]{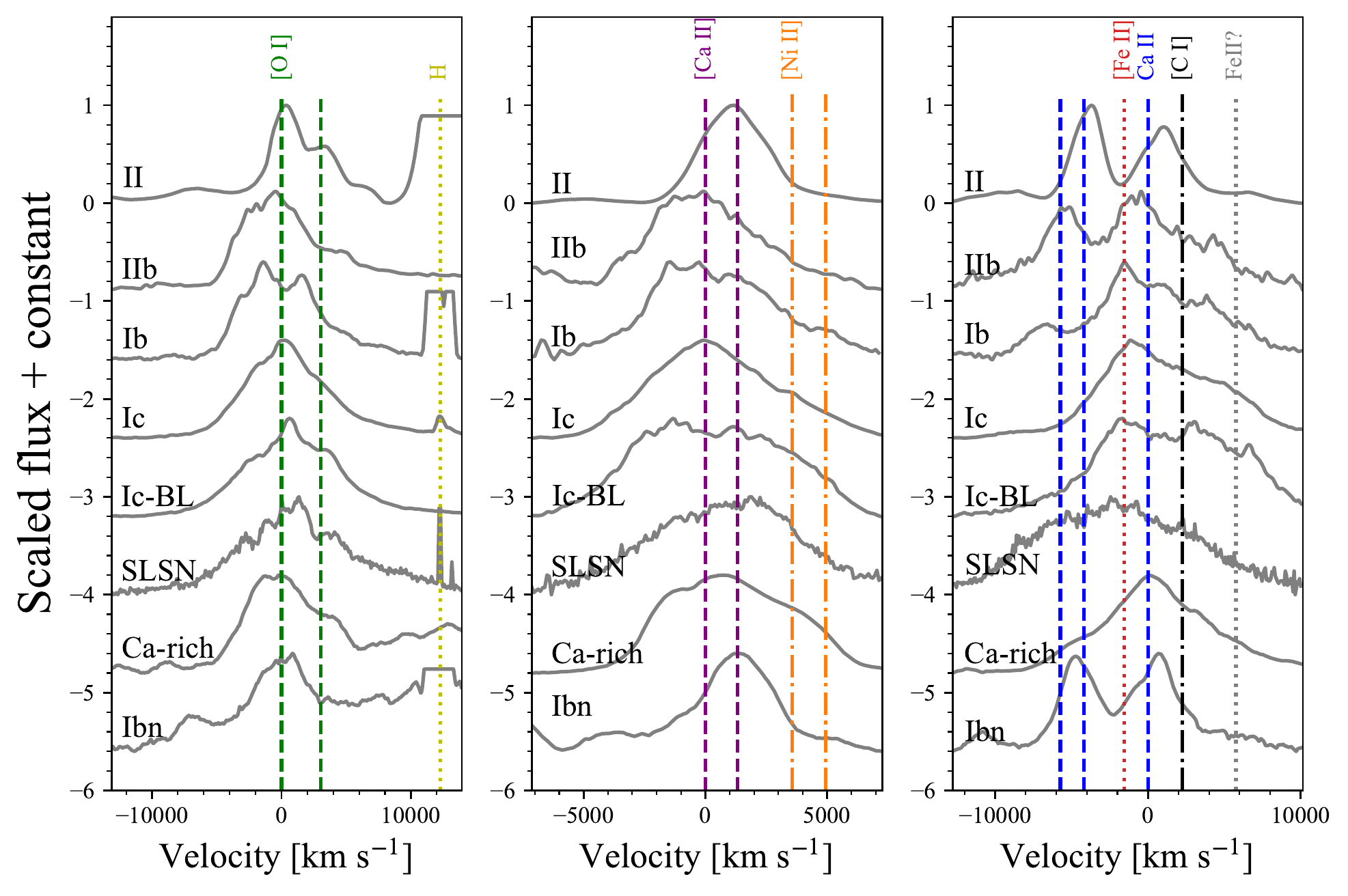}
    \caption{The three regions of interest plotted in velocity space centred on 6300 \AA\ (left), 7291 \AA\ (centre), and 8662 \AA\ (right) showing a representative SN from each subclass to demonstrate the difference in line shape. The primary sources of emission are denoted with dashed lines (\Oneb\ 6300, 6364 \AA, \caiif\ 7291, 7323 \AA, \CaII\ 8498, 8542, 8662 \AA)  with other identified sources of emission shown in dotted ( \Ha\ 6563 \AA, [\FeII] 8617 \AA, possible \FeII\ 8830 \AA)  and dash-dotted ([\NiII] 7378, 7412 \AA, [\CI] 8727). Strong \Ha\ emission has been truncated.   }
    \label{fig:pre}
\end{figure*}

The first step in the analysis was to qualitatively characterise the line emission in the various SN types to aid the more in-depth line-fitting analysis described below. 
Figure~\ref{fig:pre} shows the regions around 6300, 7291, and 8662 \AA\ in velocity space, corresponding to \Oneb, \caiif, and \CaII\ NIR features, respectively, for SNe II, IIb, Ib, Ic, Ic-BL, SLSN, Ca-rich, and Ibn. 

For the \Oneb\ 6300 \AA\ region (left panel), the SN II \Oneb\ line sits upon a continuum until rather late  ($\sim$1 year) in its evolution. The main peak is typically narrow (with velocity $500-2000$ \kms) but also includes a weaker broad component around $2000- 5000$ \kms. 
On the redward side, \Ha\ is strong in emission, and while this does not appear to contribute to the flux of the \Oneb\ line, the absorption associated with it does \citep[scattering from the H envelope still occurs to several hundred days in SNe II, e.g.][]{Jerkstrand2015}. For the remaining SNe, the \Oneb\ lines are broader 
 (typically $>5000$ \kms)
and the peaks are unimodal or bimodal in most cases. A shift to the blue of around $>1000$ \kms\ is often observed, especially in SN~Ibc \citep{2009MNRAS.397..677T}.
SNe IIb often have an emission shoulder on the red wing extending beyond 6600 \AA\ that is of unknown origin, but has been interpreted as both emission from \Ha\ or N in a He shell \citep[see][]{Jerkstrand2015,Prentice2017}. Occasionally, this `shoulder' is observed in other SN types too \citep{Prentice2017}.
In the nebular phase, the H-poor SNe lack \Ha\ emission, but narrow emission from host \Ha\ may be present. The broadness of the lines means that the \Oneb\ emission can overlap with this.

The centre panel of Fig.~\ref{fig:pre} shows the region around 7921 \AA, a region dominated by flux from \caiiflam. While this doublet is the strongest feature, there is a contribution from emission from a series of weak [\NiII] and [\FeII] lines in the vicinity, in particular [\FeII] \lam7155, as well as \HeI\ \lam7065 \citep{2015MNRAS.448.2482J}.
A flux excess on the red wing in particular can be attributed to [\FeII] \lam7388 and [\NiII] \lam7378, \lam7412. 
The emission in this region is mostly centred on 7291 \AA\ but in some objects there is evidence for a shift to the red or a peak centred on 7323 \AA\ (see the SNe Ibn and SNe II in Fig.~\ref{fig:pre}).

Finally, the \CaII\ NIR triplet region displays the most complexity of the three spectral regions under investigation. For SNe II, IIb, and Ibn, the triplet is separated into two distinct peaks and occasionally three for the low-velocity Type II where it is easier to disentangle the contributing lines. For the rest of the subtypes, the \CaII\ triplet lines are blended.
The \CaII\ NIR line remains optically thick even at these late phases \citep{Dessart2011}. 
This provides an opportunity for self-scattering if the difference in velocity between line-forming regions is such that the Doppler-shifted components overlap. From Fig.~\ref{fig:pre} it is apparent that a wide variety of profiles are present.

In the \CaII\ NIR region, most objects in our sample show a flux excess on the red side of the feature. This is suggested to come from [\CI] \lam8727  and [\FeII] \lam8617 \citep{Mazzali2010}.
The former has been shown to contribute most of the emission in this region for SNe IIb (and presumably SNe Ib) at a few hundred days post maximum light \citep{Jerkstrand2015}. 
If we consider the positions of the peaks, it can be seen that SE-SNe and SLSNe have a propensity to peak around 8617 \AA, the position of the \FeII\ line.
Another \FeII line, centred on 8830 \AA, may also be present as the emission in the SE-SNe extends out to this region and is quite strong in the SNe Ic and SNe Ic-BL (although not in the SLSN-I shown in Fig.~\ref{fig:pre}).

\section{Method for fitting line}\label{sec:method}

\begin{figure*}
    \centering
    \includegraphics[scale=0.29]{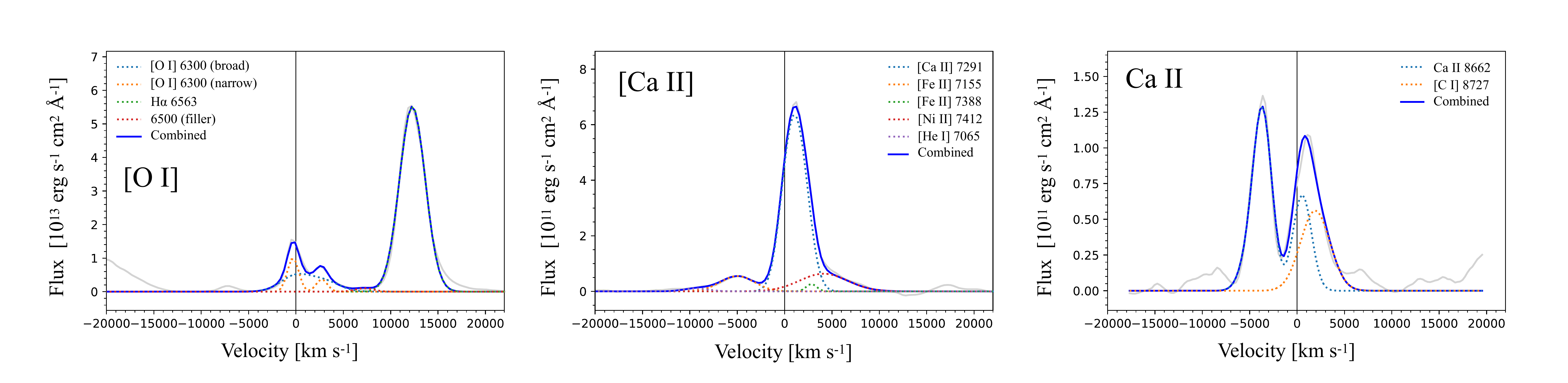}
    \caption{A demonstration of the line-fitting process for the three regions of interest  (6300, 7291, and 8662 \AA) shown in velocity space with the data shown as grey solid line and the combined fit shown as a solid blue line.   The panels are labelled with the reference components of the features contributing to that region (e.g., \Oneb\ \lam 6300 is the reference for \Oneblam). The fits consist of multiple components to account for the known emission lines in the region plus some `filler' lines to capture unidentified excess flux. {Left:} This is centred on the region around 6300 \AA, which is dominated by \Oneb. For this event, two \Oneb\ emission components are required, narrow (orange dotted line) and broad (blue dotted line) emission features. A further line is included for \Ha\  at $\sim12000$ \kms, which is largely independent in this example. A weak filler line centred at 6500 \AA\ ($\sim7000$ \kms) is also included. Middle: The \caiif\ doublet is the focus around 7300 \AA\ (\caiif\ \lam 7291, \lam7323). This is a complicated region for emission so contributions from other weaker features, such as \HeI\ \lam7065, [\FeII] \lam7155 \lam7388,  [\NiII] \lam7378, \lam7412 must be accounted for (see text for further details). {Right:} The \CaII\ triplet, centred on 8662 \AA\ and including \lam8498 and \lam8542 emission is shown in green. A contribution from [\CI] \lam8727 is also included (orange dotted line) to give the total flux (blue dotted line).  }
    \label{fig:model_fits}
\end{figure*}

The emission lines of \Oneb, \caiif, and \CaII\ NIR in our sample of 88 SNe with nebular-phase spectra were investigated using a Gaussian-fitting analysis. The primary aim of this work was to compare line-flux ratios and FWHM velocities for the chosen emission lines.
The continuum around the three features of interest (\Oneb, \caiif, and \CaII), which was to be subtracted from the emission flux, was approximated as a function of wavelength by a linear fit to the flux, with the boundaries manually defined based on the lowest continuum flux level either side of the feature. 
Care was taken to avoid absorption lines that could still be present in residual photospheric continuum.
The boundaries were then allowed to vary by the order of $\pm$ few hundred \kms\ (exact range defined per SN to avoid obvious features) to measure the uncertainty of this continuum fitting and the impact that this has on the measured fit parameters. 

In the specific case of SNe Ibn, the uncertainties on the fitting appear to be underestimated when using this method. This was due to the difficulty in not only fitting the continuum but also the contribution of permitted absorption and emission features that are still present in the ejecta, i.e.~the spectra are not fully nebular since there is continuing contribution from circumstellar interaction. A reliable and consistent method of estimating the additional uncertainties due this was not possible so we caution that the uncertainties for the SNe Ibn are likely underestimated and that there is significant scatter in the line-flux ratios for some events at early times.

The next step was to estimate the flux contributing in the wavelength regions covered by the features. As introduced in Section~\ref{sec:pre} and detailed for the individual wavelength regions below, the three lines under consideration consist of two doublets (\Oneb\ and \caiif) and a triplet (\CaII\ NIR), as well as being contaminated by the presence of other lines. Therefore, a simple approach of integrating the emission-line flux after subtracting the continuum was not appropriate. We have approximated the emission lines contributing to the three wavelength regions of interest as the sum of several Gaussian profiles.

An additional complication is that there may be multiple sources of emission from a particular transition \citep[e.g., narrow and broad components as discussed in][]{2009MNRAS.397..677T}. We discuss the use of single or multiple sources of emission for a particular feature below in Sections \ref{fit6300} (\Oneb\ region), \ref{fit7300} (\caiif\ region) and \ref{fit8600} (\CaII\ NIR region).

We now describe the generic-fitting method for including Gaussian profiles. We have modelled the emission from the constituent transitions of each multiplet (e.g.~6300 \AA\ and 6364 \AA\ in \Oneb) as individual Gaussian profiles, with the total line emission represented by a sum of these Gaussian profiles. 
For each multiplet, one transition was selected to be the reference and defined by three free parameters: intensity, width, and offset of the peak from the rest wavelength. The width and offset of the peak were fixed as the same for the other lines of the doublet or triplet being studied. The intensity was allowed to vary within the range allowed for optically thick and thin regimes, with the exact values used for each feature and region detailed below. In most SNe, the offset of the peak wavelength was limited to $\pm{1000}$ \kms\ as this broadly covers any potential shifts due to the spectral resolution of the instrument\footnote{
We note that spectral resolution can affect the measured widths if the resolution is comparable to the feature width. The sample used here contains spectra with a range of resolutions. The lowest resolution spectra in the sample are from the LT and the SPRAT spectrograph (resolution of $\sim 850$ \kms). This is smaller than any measured width in our sample, which negates the need to convolve the spectra with a line-spread function \cite[see discussion in][]{2012MNRAS.420.3451M}.},   as well as the intrinsic scatter of the peak wavelength position of the sample. In some specific cases, a larger shift in the central position of the Gaussian was required, e.g.~the line shift of the \Oneb\ line in SNe Ic can be as much as $\sim1300$ \kms\ \citep[e.g.][]{2009MNRAS.397..677T,2022arXiv220111467F} and for these events, the allowed peak offset region was extended. This was applied on a case-by-case basis by checking any outputted fits that have values (e.g.~width or offset from the peak) that are at the input boundary limits and performing visual inspection.

As an example, to construct the \Oneb\ doublet profile, the reference line was taken to be the transition at 6300 \AA. The three parameters given above were defined for this transition and the width and offset of the peak were fixed to be the same for the other line of the doublet at 6364 \AA. The intensity of the \Oneb\ \lam6364 was allowed to vary from 0.3 -- 1 times the peak intensity of the \lam6300, as discussed in Section \ref{fit6300}. The initial parameters for each component were randomly initiated within the allowed priors, then a Metropolis-Hastings algorithm was used to search the parameter space for the best overall fit. To obtain an estimate of their uncertainties, 10\,000 runs per fit were undertaken.
Below we describe the constraints used to fit the flux in the three regions of interest, with example fits to each line shown in Fig.~\ref{fig:model_fits}.

\subsubsection{The \Oneb\ 6300 \AA\ region}
\label{fit6300}
For \Oneb\ \lam6364, the intensity was limited to 0.25 -- 1 times that of \Oneb\ \lam6300 following \citet{2009MNRAS.397..677T} and is related to the density of the ejecta in the region where the line formation occurs. A 3/1 ratio suggests an optically-thin ejecta. Following the results of \citet{2009MNRAS.397..677T}, two sources of \Oneb\ emission were used by default, one narrow and one broad (see the left-hand panel in Fig.~\ref{fig:model_fits}). These distinct contributions were required to account for previously identified asymmetries in the ejecta that manifest as multiple components in the spectra and were needed in $\sim$80 per cent of SNe studied. In $< 20$ per cent, a third very weak component was also included but did not affect the fits. Most Type II SNe at later phases can be equally well fit with a single \Oneb\ component but the fit results are consistent within the uncertainties. This is because the first component dominates in intensity over the second component. 

A Gaussian representing \Ha\ was also included in this region; it was chosen as broad (up to 8000 \kms) for SNe II and narrow (up to 100 \kms) for everything else (representing host galaxy contamination). Additionally, a broad Gaussian was included between 6400 and 6500 \AA\ in order to account for the `shoulder' emission seen in many objects, especially SNe Ib/IIb \citep[see][]{Prentice2017}. While we used a Gaussian profile for this feature, others such as the Type II analysis of \citet[][]{2016MNRAS.462..137T} have used a flat-topped profile. The difference gives small variations in the results but the crucial point is to capture this excess flux and its contribution to the total flux in the \Oneb\ region. Without it, the \Oneb\ flux is greatly overestimated and is one of the main problems associated with simply integrating the flux in this region. 

\subsubsection{The \caiif\ 7300 \AA\ region}
\label{fit7300}
For the \caiif\ \lam7323 feature,  the peak intensity was allowed to vary between 1--10 of the 7291 \AA\ component \citep{NIST_ASD}. Examination of narrow-lined SNe II 2008bk and 2005cs suggested that a similar range would be valid \citep{2012MNRAS.420.3451M}, while other studies have used values of unity \citep[][]{2016MNRAS.462..137T,Jerkstrand2017hb}. A single source of the \caiif\ doublet was used throughout, e.g.~different broad and narrow components were not included. 
In addition to the \caiif\ doublet, we also included emission from \HeI\ \lam7065\ and the weak forbidden \FeII\ and \NiII\ lines referenced in Section~\ref{sec:pre} in order to account for additional flux at the base of this feature.

\subsubsection{The \CaII\ NIR region}
\label{fit8600}
Finally, we investigated the \CaII\ NIR triplet region. Here, the \CaII\ 8662 \AA\ line was used as reference and the intensity of both the 8498 \AA\ and 8542 \AA\ components were allowed to vary between 1.1 and 1/50 that of the reference 8662 \AA\ line. This ratio is poorly constrained and suffers from scattering effects \citep{Dessart2011}. A single component of the \CaII\ NIR triplet was used throughout.
We also included the [\CI] \lam8727 line in every instance because this greatly improved the fit to the red wing. The [\CI] line is also clearly present as a distinct feature in some of the narrow-lined SNe II and becomes more prominent over time in the SNe IIb.
\FeII \lam8830 was included in some of the SNe Ic and SNe Ic-BL to account for the far red wing in these objects, as they were fit insufficiently well with [\CI] alone.

A difficulty in this region is that fitting these lines to the feature in the SNe Ic resulted in poor fits unless the \CaII\ offset was allowed to centre on the peak of the emission around 8600 \AA, the approximate position of an additional [\FeII] line at 8617 \AA\ (see Fig.~\ref{fig:pre}). {We tested the inclusion of this additional [\FeII] \lam8617 feature in SNe Ic but the result of this was the removal of any significant component of \CaII\ emission, leaving the feature a mix of just [\FeII] and [\CI]. Restricting the width of the [\FeII] emission met with poor fits and there was also no physical reason to do this. } {Similarly, a poor fit was obtained if the entire emission is assumed to be solely from \CaII, with no [\CI] emission.  } The conclusion is that a proper investigation of the contribution of each feature to the emission in SNe Ic requires more complex techniques than simple Gaussian fitting and is beyond the scope of this work. To provide an estimate of the properties of the \CaII\ NIR emission in SNe Ic, we returned to the original approach of fitting the feature with a combination of \CaII\ and [\CI] as this represents an average of the two extremes. The difference in integrated flux between this approach and assuming that the entire emission is \CaII\ is $\sim35$ percent.

\section{Results}\label{sec:results}

In this section, we present the results of measuring the flux ratios as a function of time for the different emission lines studied, the flux ratios in terms of physical parameters, such as ejecta mass and kinetic energy, as well as a comparison of the FWHM velocities for the sample.

\subsection{Emission line flux ratios} 
\begin{figure*}
    \centering
    \includegraphics{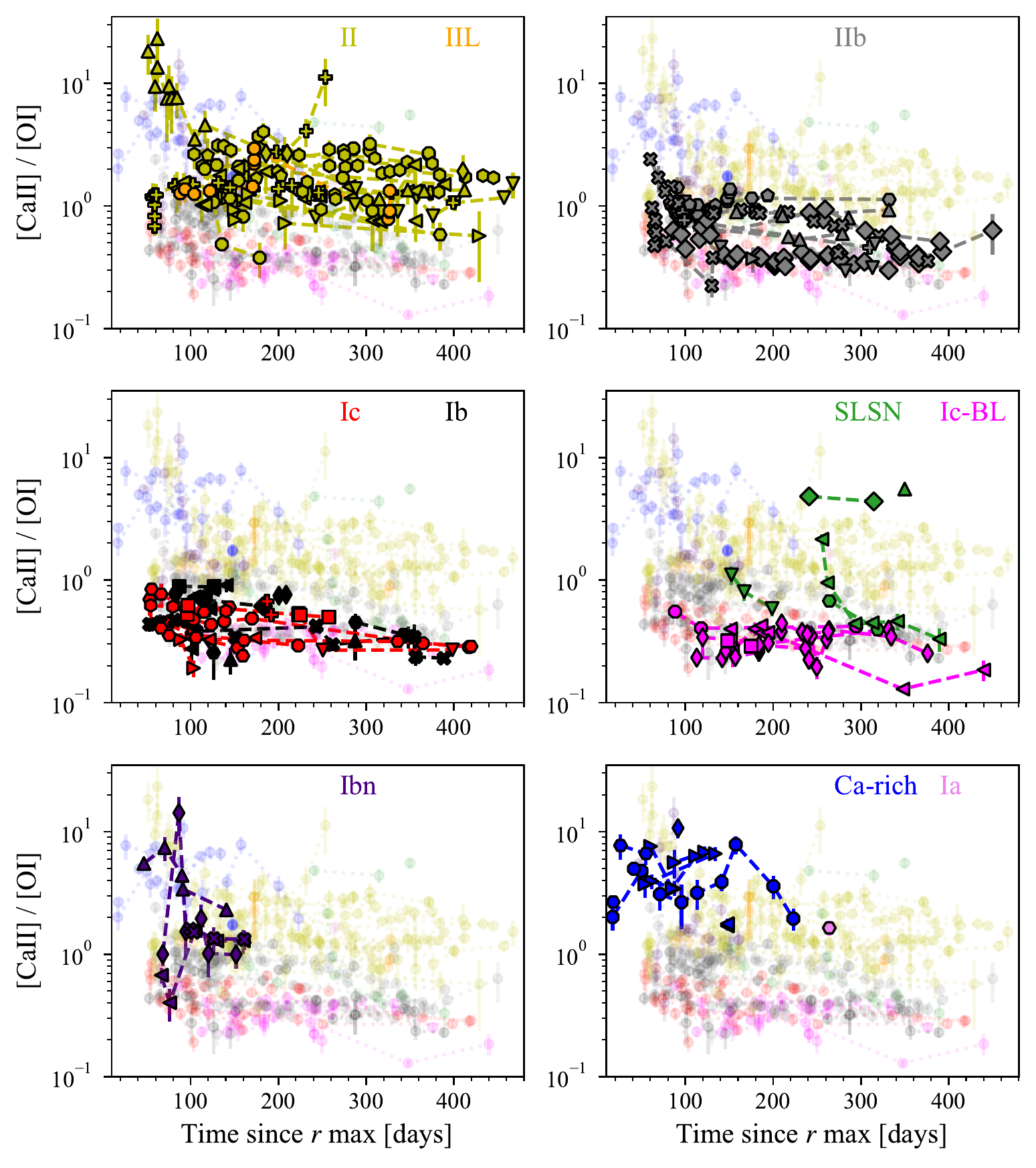}
    \caption{The ratio of \caiif\ \lam\lam7291, 7323 to \Oneb\ \lam\lam6300, 6364 for the various SN types, highlighted by sequence. Each panel shows the different labelled SN type in non-transparent points, with the rest of the sample as semi-transparent points for comparison. A rapidly changing ratio before 100 d is indicative of the transition into the early nebular phase.  The SLSNe evolve on a much longer timescale than the other events and this is reflected in their ratio curves. In the middle right panel, for the SLSNe, \sn2010md is shown by green diamonds and \sn2017egm by the upward green triangle. As discussed in Section \ref{sec:method}, the uncertainties on the Ibn events are underestimated due the difficulty in removing the contributions from remaining absorption features from the circumstellar interaction.     }
    \label{fig:caiif-O}
\end{figure*}

\begin{figure*}
    \centering
    \includegraphics{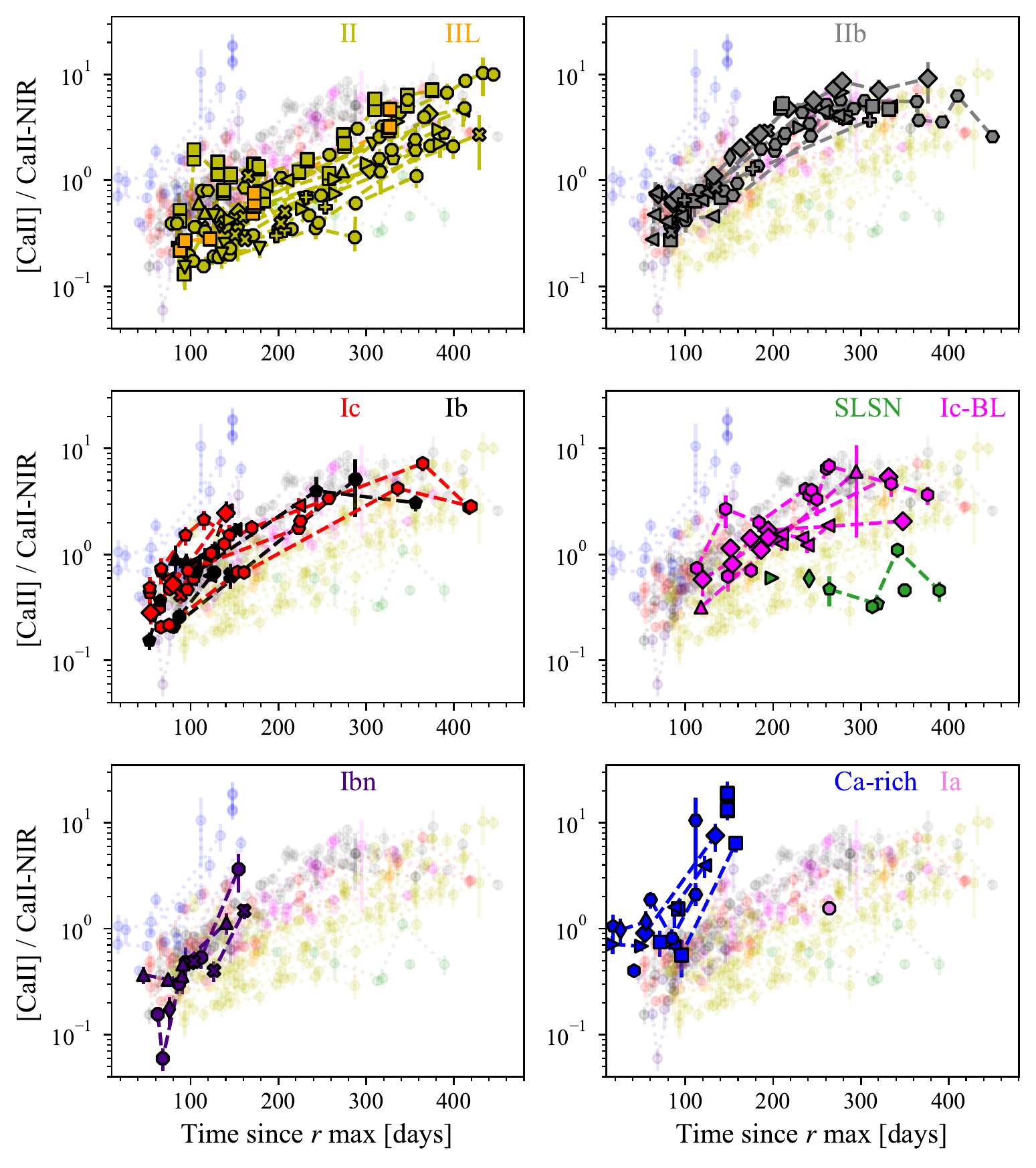}
    \caption{The evolution of the \caiif\ \lam\lam7291, 7323 to \CaII\ NIR ratio for different SN types. The same breakdown of types is shown in each panel as in Fig.~\ref{fig:caiif-O}.} 
    \label{fig:caiif-canir}
\end{figure*}

\begin{figure*}
    \centering
    \includegraphics{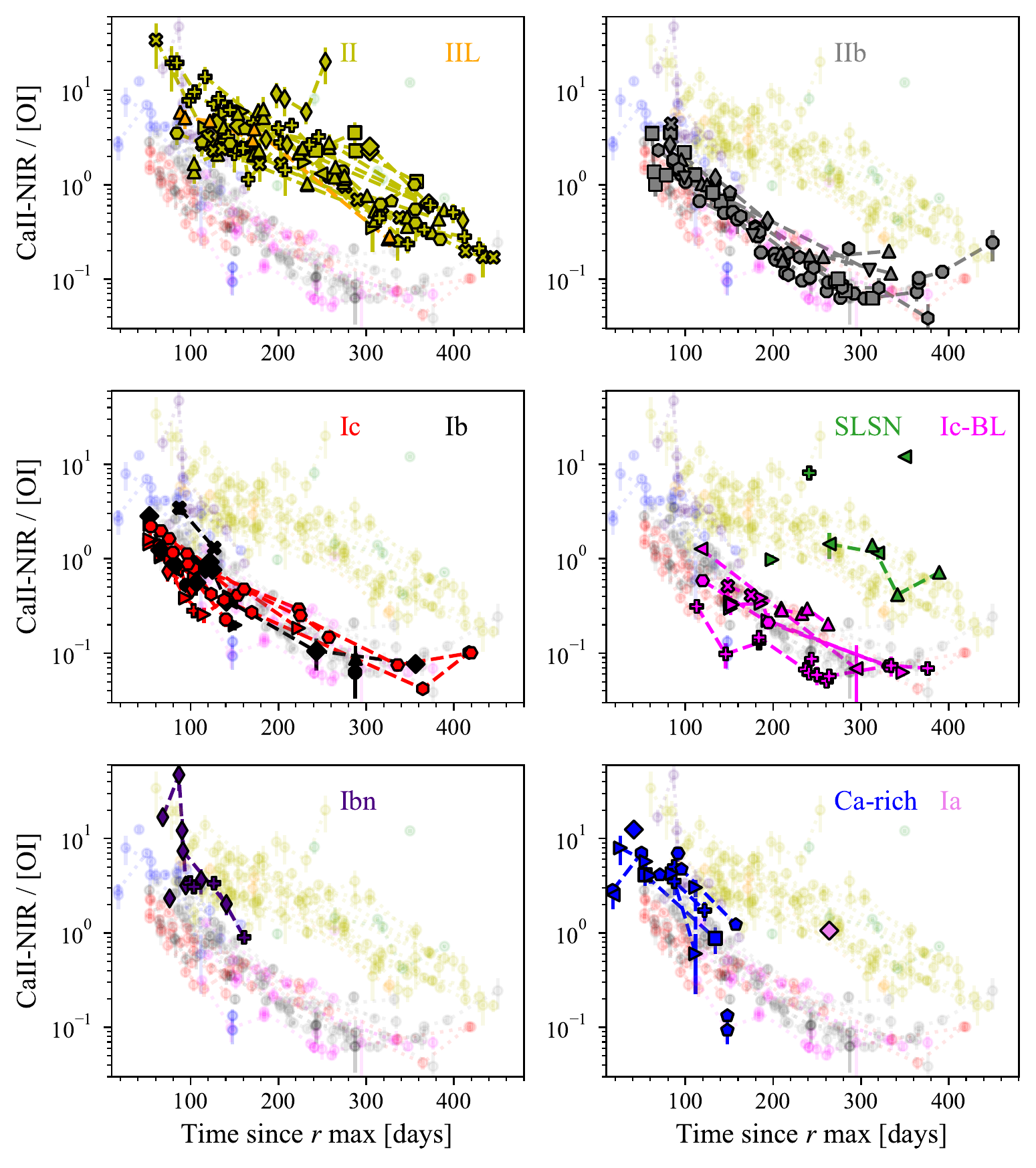}
    \caption{The tracks of different SN types for the ratio \CaII\ NIR to \Oneb\ \lam\lam6300, 6364.  The same breakdown of types is shown in each panel as in Fig.~\ref{fig:caiif-O}.} 
    \label{fig:canir-O}
\end{figure*}

We wished to investigate how the strength of emission features vary across SN types, as well as look for similarities. Since absolute flux calibration is prone to large uncertainties, particularly due to the sparse photometric data at these epochs, we use line ratios throughout.  The results for the line ratios as a function of time are plotted in Fig.~\ref{fig:caiif-O} for \fo, Fig.~\ref{fig:caiif-canir} for \fc, and Fig.~\ref{fig:canir-O} for \co. Some general trends are noticeable;  \fo\ is effectively constant in time for most transient types (a rapid early decline is sometimes seen and this represents the transition from photospheric to early-nebular phase). A decreasing trend for \fo\ ratios is seen in the order of SNe II, followed by IIb, Ibc and finally Ic-BL with the smallest ratios. This is in agreement with the results of \citet{2022arXiv220111467F}, where a similar overall trend in the \fo\ ratio for SE-SNe was seen. The interacting SNe Ibn and Ca-rich transients are observed, on average, at earlier times than the rest of the sample and have generally higher \fo\ ratios with a larger scatter between objects. The SLSNe in our sample have a range of \fo\ ratios, which generally overlap with the region occupied by SNe II but where multiple epochs are available they appear to show a slower evolution (see Section \ref{sec:slsne} for further discussion of this).

As shown in Fig.~\ref{fig:caiif-canir}, the \fc\ flux ratio increases with time and in a linear way in log space. 
The \co\ ratio decreases with time, with varying slope depending on the SN type. The SNe II and the SE-SNe are clearly separated in this parameter space (see Fig.~\ref{fig:canir-O}). For both \fc\ and \co\ line ratios, the SE-SNe show a change in the slope after 300 d, suggesting that this break is driven by the \CaII\ NIR line. The SNe II do not show this break but this may be a consequence of their slower time evolution and so the break may occur beyond the phase covered by observations.
In the following sections, we describe the evolution and relative position of the three line-flux ratios for each class of transients studied.

\subsubsection{SNe II}

With their lower kinetic energies and narrower lines, SNe II are the easiest objects on which to perform this kind of analysis. In most cases, the individual lines are distinct and unblended, allowing a well-constrained fit to each feature.
SNe II tend to enter the nebular phase at a later time than other objects, and their spectra are dominated by strong \Ha\ emission. It is found that SNe II have \fo\ $= 0.7 -3$, with a typical  (median) value of $\sim 2$ from about 100 d past maximum (Fig.~\ref{fig:caiif-O}). This is in excellent agreement with the \fo\ flux ratio values obtained by \cite{2012MNRAS.420.3451M} for a smaller sample of Type II SNe.
In the \fc\ plane shown in Fig.~\ref{fig:caiif-canir}, SNe II follow a linear track with increasing ratio with increasing phase, and in \co\ (Fig.~\ref{fig:canir-O}) they follow a decreasing linear track with increasing time. 
The spectroscopic epoch of each event is scaled relative to maximum light in the $r/R$ photometric band. For \sn1987A-like events this can be $40+$ days after explosion while for normal SN~II this can be a few days. Interestingly, we find that this approach results in the events following similar tracks rather than having multiple ratio-curves shifted by tens of days.

\subsubsection{SNe Ic and Ib}
SNe Ic and Ib are found to have \fo\ ratios of  0.2 -- 1, with the SNe Ic-BL having marginally lower ratios of \fo\ $= 0.2 - 0.5$. 
Typically, the SNe Ibc are found to lie in a distinct region to the SNe II, with 
lower \fo\ and \co\ ratios and higher \fc\ ratios. There is some small overlap between the two groups in the \fo\ plane. However, the two subtypes are completely separated in the \fc\ plane.
The SNe Ibc also have a steeper gradient with time to the tracks they follow in \fc. The tracks of the SNe Ic are slightly shallower than those of the other SE-SNe. 
Turnovers can be observed in the \fc\ and \co\ ratio curves between 250 -- 300 and 300 -- 400 d, respectively.
Such a change is not seen in the SNe II out to 500 d.

\subsubsection{SNe IIb}
The SNe IIb have values of the \fo\ flux ratio of 0.3 -- 2 once they have settled into the nebular phase and any contribution from photospheric features become sub-dominant. This places them partially overlapping the region of the H-poor SE-SNe but with several objects having larger \fo\ values,   more similar to those seen for SNe II. 
The light curves of SNe IIb are known to fall into two classes, those that show a prominent shock cooling tail (extended) and those that do not \citep[compact; e.g.,][]{2012ApJ...752...78S}. 
We find that the SNe IIb with the largest ratios at $>150$ d are mostly from those defined as belonging to the compact class, but there is not enough data at this time to draw any firm conclusions from this.

\subsubsection{SLSNe}
\label{sec:slsne}
The SLSN subset studied in this work are all of the SLSN-I type, meaning that they lack signatures of H in their spectra.
The SLSN sample is small with just six objects but the majority appear to show a similar evolution in their line ratios to the SNe IIb and SNe II, in that the ratio decreases over time before levelling off. However, in a notable difference to the SNe IIb and II, this levelling off occurs a few hundred days later in the SLSNe. 
This is consistent with their slower photometric evolution in comparison to `normal' CC-SNe \citep[e.g.][]{2019ARA&A..57..305G}.

Two exceptions are seen in SNe 2010md and 2017egm, which displayed the largest \fo\ values seen for any SNe in the sample (\fo\ $= 5-7$) at a time when other SLSNe have similar ratios to the H-rich CC-SNe (marked as an upward green triangle and green diamonds in Fig.~\ref{fig:caiif-O}, respectively).
\sn2010md is a rare example of a SLSN-IIb (demonstrating that some of its H envelope was intact at the time of explosion), and was noted by \citet{2018ApJ...855....2Q} to be an outlier to the existing SLSN dataset.  
Examination of its spectra show that the \caiif\ lines are strong, but \Oneb\ is very weak and blended as evidenced by its large \fo\ ratio. The spectra appear to still be somewhat photospheric.
Likewise, \sn2017egm was also strong in \CaII\ emission and weak in \Oneb. However, in this case the \Oneb\ line is distinctive and appears to be sitting upon a blue continuum. Owing to their relatively high redshift, few SLSNe have spectra that extend far enough to observe the \CaII\ NIR line. 
We do obtain some measurements, however, and as before the ratios are consistent with the other SN types if the much longer evolutionary timescales (a factor of a few) for SLSNe is taken into account.

\subsubsection{SNe Ibn}
The spectra of SNe Ibn are dominated by lines of \HeI\ and \FeII\ as a consequence of ejecta interaction with CSM. Consequently, these lines can mask SN line emission if the source of emission is located within the radius of interaction \citep[see][]{2020MNRAS.499.1450P}. The SN Ibn sample is small (four events) since few SNe Ibn are observed longer than 60 d after maximum light - well before CC-SNe enter the nebular phase - and so most strongly interacting SNe are not observed to display emission from \Oneb\ \lam\lam6300,6364, \caiiflam, nor \CaII\ NIR.  As discussed in Section \ref{sec:method}, we also note that the uncertainties on the SN Ibn flux measurements are likely underestimated due to the additional complications caused by the contributions from features due to circumstellar interaction.
Analysis of the few objects that did display these lines shows that they are located in a similar position to SNe II in the \fo\ and \co\ planes.
When considering just the Ca lines, however, SNe Ibn follow the same track as the SE-SNe.

\subsubsection{Ca-rich transients}
The ratio \fo\ of Ca-rich transients is well studied in the literature \citep[e.g.,][]{Valenti2014}. 
Here the results of the previous studies are replicated, with these objects taking values of \fo\ $= 1-10$. In the \fc\ plane the Ca-rich transients have larger ratios than any other objects and display a much steeper gradient in the tracks they follow.
Despite having extreme values in the previous ratios, when \co\ is considered the Ca-rich objects sit between the SE-SNe and SNe II events. Although the data is lacking due to their rapid evolution, the results suggest that they have a steep gradient to this track as well, which crosses the SE-SN track.

Ca-rich events are characterised by fast-evolving light curves \citep{Lunnan2017} and an early evolution into the nebular phase, which is dominated by emission from the \caiif\ and \CaII\ NIR lines of interest in this work.
These objects are typically observed for short periods of time compared with other events due to their rapid evolution and the majority of events do not have observations past 100 d. 
Ca-rich events also likely consist of different populations with potentially both SN Ia and CC-SN origins \citep[e.g.][]{2017ApJ...846...50M, 2020ApJ...905...58D} but this is not apparent in the relative ratios between the objects. 
It is clear, however, that they sit apart from other H-poor transients.

\begin{figure}
    \centering
    \includegraphics[scale=0.47]{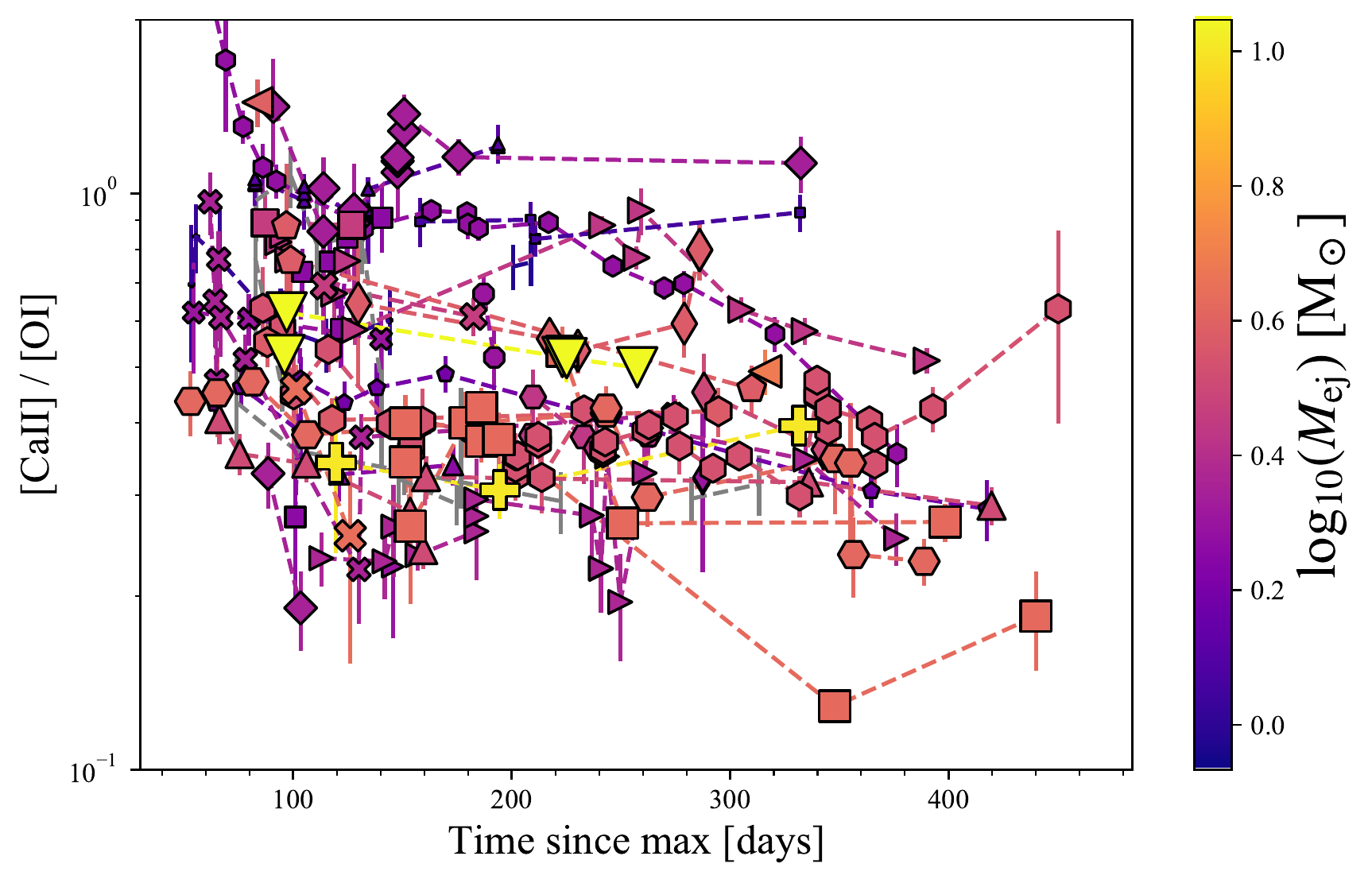}
    \includegraphics[scale=0.47]{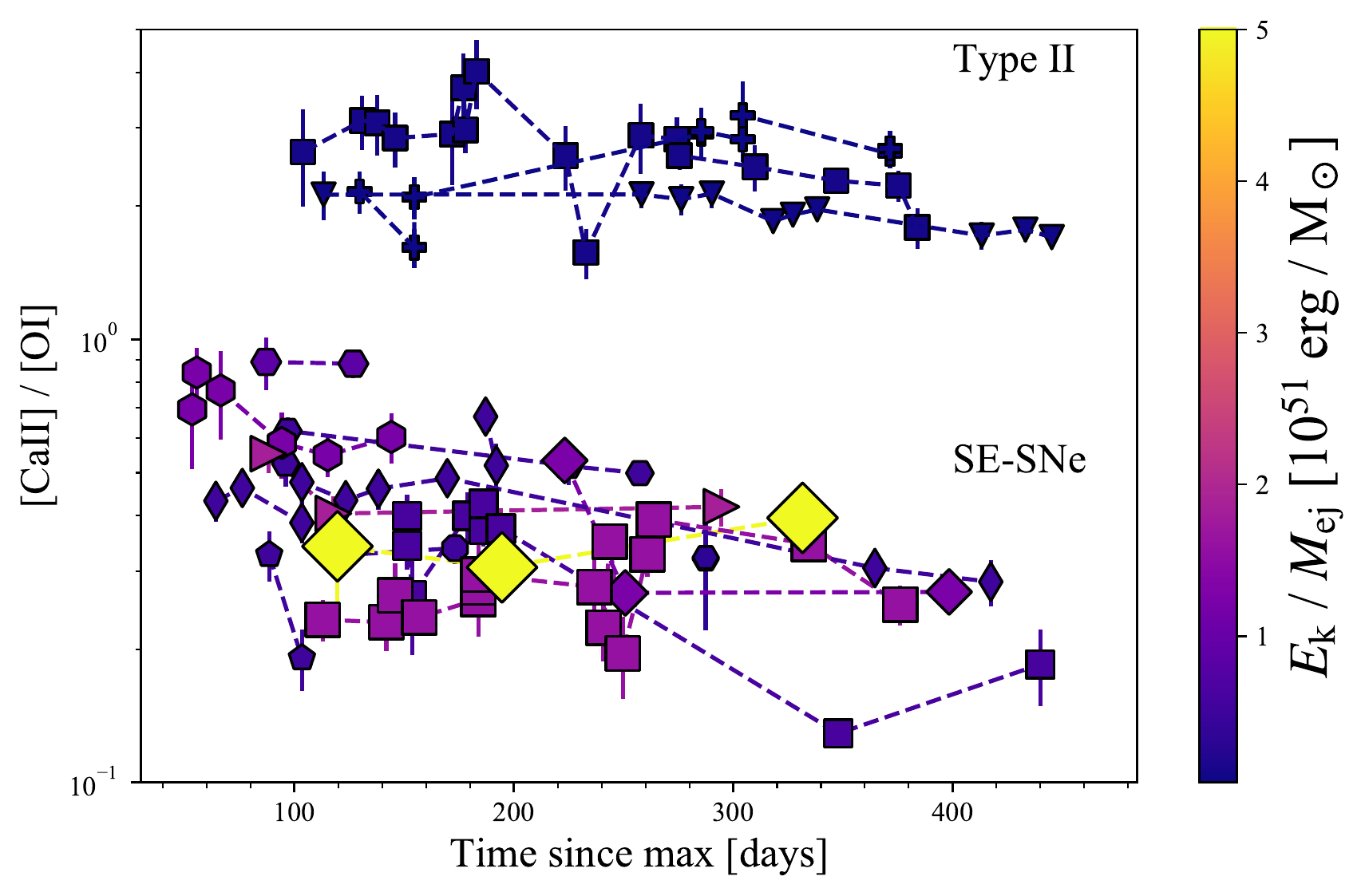}
    \caption{Upper panel: SE-SNe (of all types) \fo\ tracks as a function of time and colour-coded based on their ejecta masses as calculated from detailed light-curve modelling from the literature. The least massive objects have smaller marker sizes. The typical ejecta mass is $\sim2$ \msun, and the maximum is 10 \msun. This plot demonstrates that \fo\ is not a proxy for core O mass within SE-SN subtypes. Lower panel: The specific kinetic energy obtained from the detailed modelling for the available sample of SE-SNe and three SNe II 1987A, 1999em, 2004et. This demonstrates that the position of an object within \fo\ is not related to its \eom\ within the individual subtypes, although SE-SNe have larger \eom\ than SNe II. All references are provided in Table~\ref{tab:sample} and Section~\ref{sec:params}.  }
    \label{fig:mass-eom}
\end{figure}

\begin{figure*}
    \centering
    \includegraphics[scale=0.7]{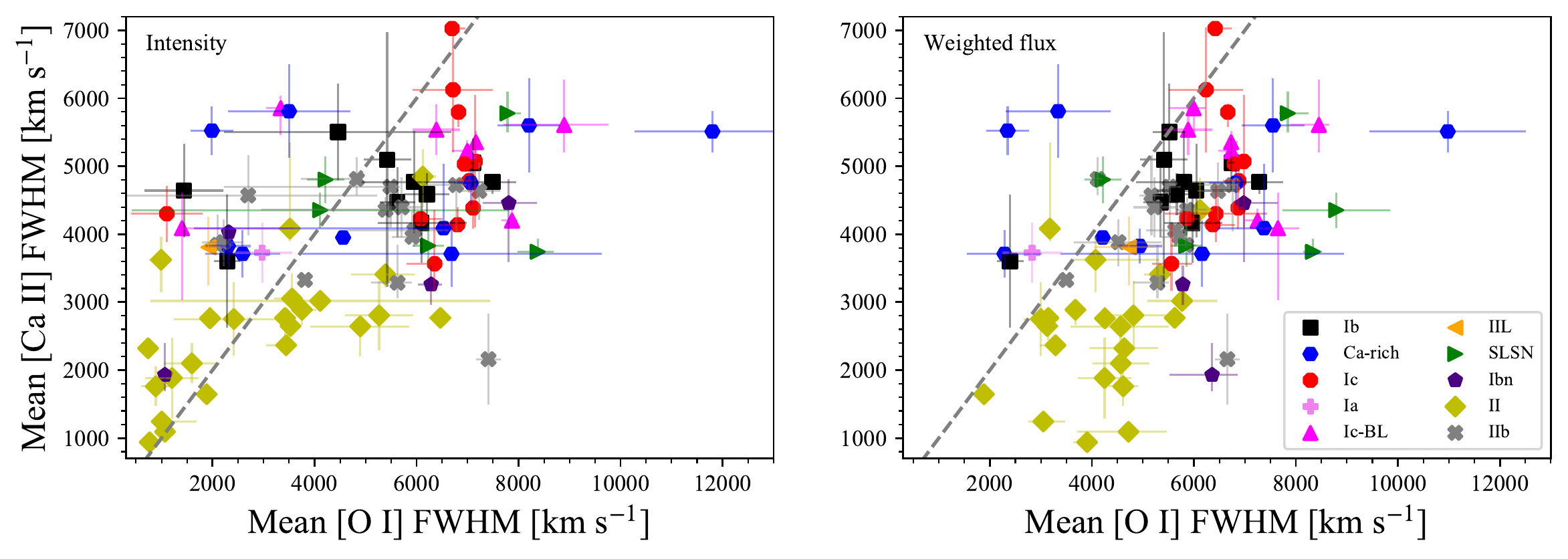}
    \caption{Time-averaged FWHM velocities of the \caiif\ \lam7291 and \Oneb\ \lam6300 Gaussian fits for the SN sample, the dashed grey line is the line of unity. Left: FWHM selected on the maximum line intensity at each spectroscopic epoch. Right: The FWHM selected by taking the average of each component weighted according its contribution to the total flux of the line per spectroscopic epoch.  }
    \label{fig:fwhms}
\end{figure*}

\subsection{Ratios in relation to physical parameters}\label{sec:params}

SNe have different physical properties depending on their type, and many of these, such as ejecta mass, \mej\ and kinetic energy, \ek, depend on the explosion mechanism and the mass of the progenitor star. 
These properties can be estimated through a variety of means, with varying degrees of accuracy.
The \ek, in particular, is hard to determine, because the majority of this energy is in the thin high-velocity ejecta that is only seen at early times and has only a small impact on light-curve shape. 
Thus, for \ek\ and \mej\ the only values that are used here are those calculated from spectral modelling, hydrodynamic codes, explosion modelling, or if the spectra and light curves of an object are well matched to one which has been analysed in this way. 
We consider the Type II SNe 1987A \citep{Blinnikov2000}, 1999em \citep{2020ApJ...902...95L}, and 2004et \citep{2004AstL...30..293U}, in addition to a sample of SE-SNe \citep{Woosley1994,Nomoto1994,Pian2006,Mazzali2008,Mazzali2009,2012MNRAS.422...70H,2013MNRAS.432.2463M,Fremling2016,Mazzali2017,2018MNRAS.478.4162P,Prentice2019,Teffs2021}. These model efforts come from a variety of codes but all are significantly more detailed than simple `Arnett' calculations. While non-negligible systematics likely exist between these modelling methods, they are still broadly compatible and worthy of comparison to the measured flux ratios.

To consider if there is any relationship between the line ratios and physical parameters, we show the \fo\ ratio as a function of \mej\ and specific kinetic energy, \eom, in Fig.~\ref{fig:mass-eom}.
The upper panel shows the \fo\ ratio as a function of phase for SE-SNe with their \mej\ from obtained from the literature displayed as a colour bar. It is evident that the SNe are not ordered by mass. 
Nebular-phase modelling demonstrates that O is the primary constituent of the ejected mass in SNe~Ic, and this increases for a more massive progenitor \citep[e.g.,][]{Jerkstrand2015}.
For the SNe~Ib/IIb, measurements of the total \mej\ are harder to make as H and He contribute little in terms of features to the nebular spectra of SE-SNe but in reality may make up 1--2 \msun\ of the ejecta \citep{Jerkstrand2015}.

The lower panel of Fig.~\ref{fig:mass-eom} shows the \fo\ curves for SE-SNe and the Type II SNe studied, with the colouring to reflect the specific kinetic energy, \eom.
The Type II SN sample has lower values for \eom\ (typically $\sim0.1$ $\times$ $10^{51}$ erg/\msun), and is offset from the SE-SNe, where \eom\ varies from $0.25 - 4$ $\times$ $10^{51}$ erg/\msun.  There appears to be no correlation between the \fo\ flux ratio and the parameters within each SN type. However, we note again the caveat that this comparison includes different codes and methods for calculating the \mej\ and \ek.

\subsection{FWHM velocities}\label{sec:fwhm}

For each best-fitting model for the \Oneb, \caiif, and \CaII\ NIR features, there is a defined FWHM velocity for \Oneb\ \lam6300 (\fwhmo), \caiif\ \lam7291 (\fwhmf), and \CaII\ \lam8662 (\fwhmc), that is shared with the other components of the fit.
There are two ways to interpret the FWHM values. The first is to assume that it defines emission from a spherical region or shell centred on the explosion. The second is to assume that it represents the spatial distribution of a blob of ejecta \citep[e.g.,][]{2009MNRAS.397..677T,2017MNRAS.469.1897S,2022arXiv220111467F} . In both scenarios, it provides a rough estimate of the distribution of material within the ejecta.

To obtain a single velocity measurement per object for comparison, we have averaged the velocities in time per object by fitting a second-order polynomial to the available measurements in the region $100-200$ d (when most objects had observations and the evolution is effectively flat). These are weighted according to the uncertainties on the measurements and interpolated to a per day value which was then averaged over the 100 d time frame. The uncertainties on this method were found using bootstrap resampling, by randomly removing points and repeating the process.  Figure~\ref{fig:fwhms} plots the time-averaged \fwhmo\ against \fwhmf\ via two methods. The first, given in the left-hand panel, selects the \Oneb\ component with the highest intensity. The second, given in the right-hand panel, takes all the \Oneb\ components e.g.~broad and narrow for a particular spectroscopic observation and returns an average of the FWHM weighted by the contribution to the total flux of each line.

The latter method of the weighted FWHM gives a better indication of the typical velocity for the distribution of \Oneb\ emission (which could be spherical or blob-like).
The former, based on peak intensity, provides a better measure for the distribution of the main energy injection, or regions of greater O abundance \citep{Teffs2021}.
In almost every case, the component with the highest intensity and the components with the largest flux are those most centrally located.
Components with offsets significantly further out are almost always weak and narrow, so contribute little to the weighted average in Fig.~\ref{fig:fwhms}.

\begin{figure}
    \centering
    \includegraphics[scale=0.6]{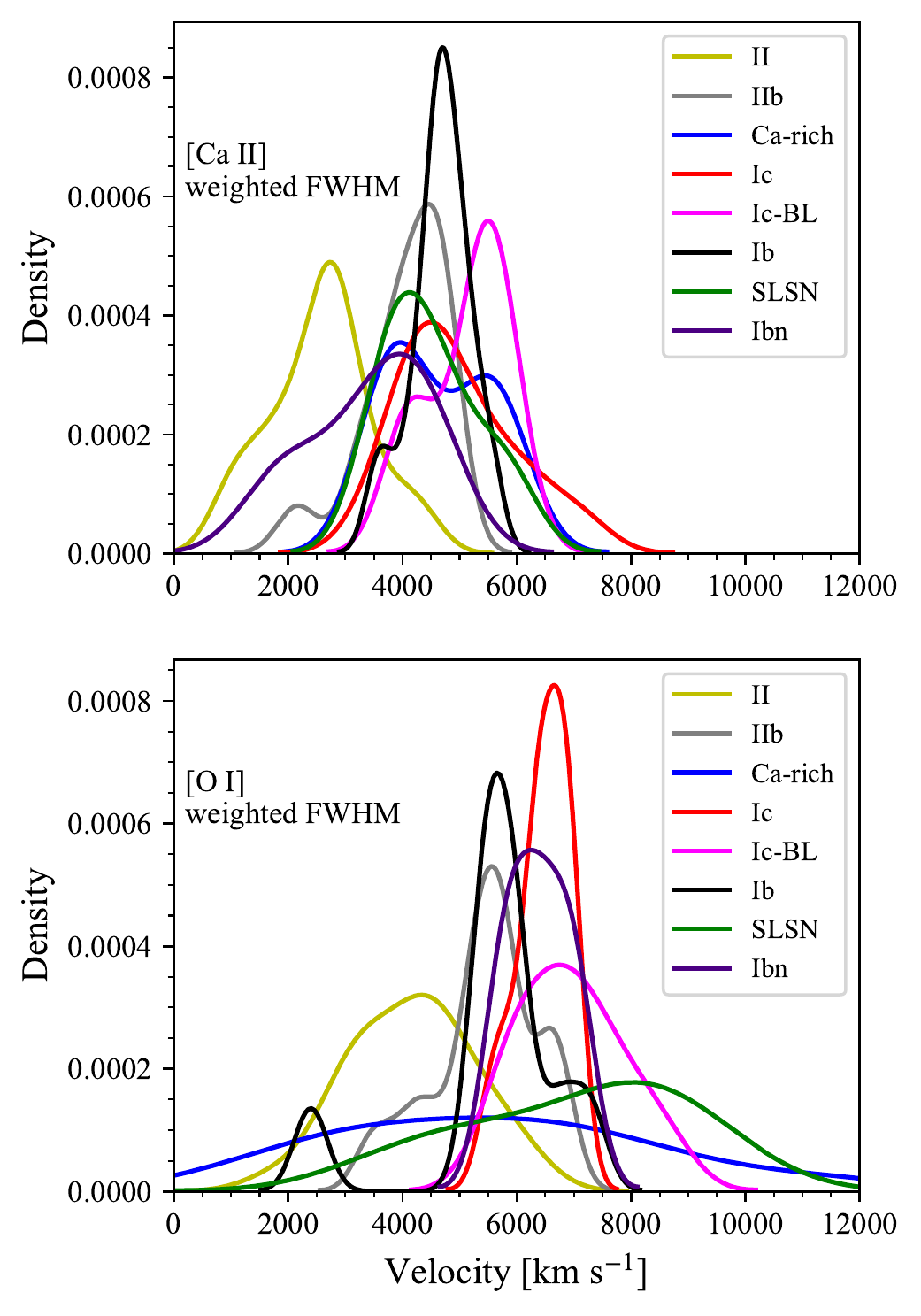}
    \caption{Upper panel: Density distributions for the time-averaged flux-weighted \caiif\ \fwhmf\ measurements. Lower panel: The same for \Oneb\ \fwhmo. The method of calculating the time-averaged values is described in Section \ref{sec:fwhm}.} 
    \label{fig:fwhm_dist}
\end{figure}

\begin{table}
    \centering
    \caption{Table of median time-averaged flux-weighted \caiif\ \fwhmf\ and \Oneb\ \fwhmo\  values for the population distributions by SN type. The uncertainties are $\pm$1 sigma on the distributions.}
    \begin{tabular}{lcc}
    \hline
    Type & FWHM$_{7921}$ & FWHM$_{6300}$ \\
     & [\kms] & [\kms] \\
     \hline
II & $2752\pm^{1049}_{479}$ & $4253\pm^{1120}_{905}$  \\
IIb & $4208\pm^{846}_{486}$ & $5579\pm^{1011}_{885}$  \\
Ib & $4705\pm^{408}_{371}$ & $5736\pm^{359}_{705}$  \\
Ic & $4784\pm^{592}_{1147}$ & $6431\pm^{346}_{446}$  \\
Ic-BL & $5355\pm^{1160}_{264}$ & $6737\pm^{746}_{947}$  \\
SLSN & $4354\pm^{556}_{803}$ & $7843\pm^{314}_{1353}$  \\
Ibn & $3644\pm^{1079}_{604}$ & $6357\pm^{391}_{423}$  \\
Ca-rich & $4422\pm^{2585}_{657}$ & $5549\pm^{2773}_{1925}$  \\
\hline
    \end{tabular}
    \label{tab:fwhm_stats}
\end{table}

\begin{figure}
    \centering
    \includegraphics[scale=0.7]{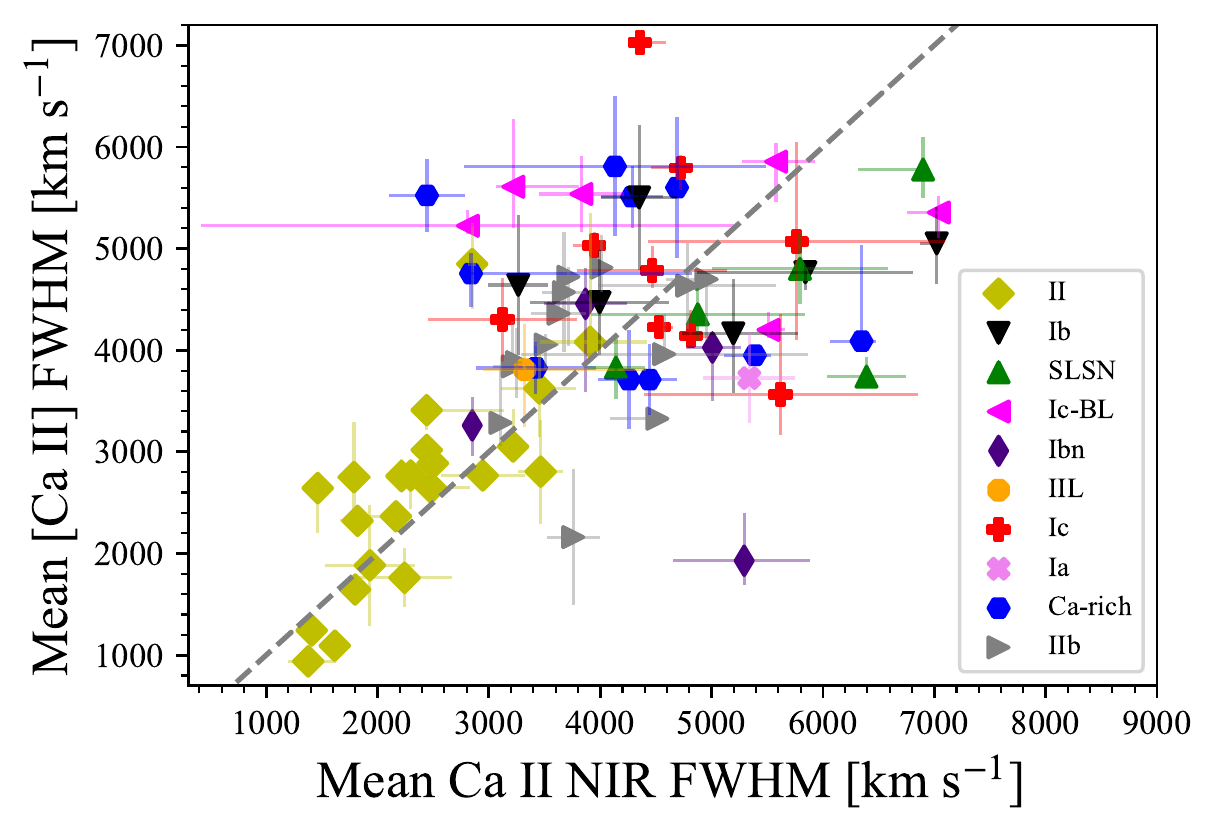}
    \caption{Time-averaged ($100-200$ d) FWHM measurements of \caiif\ and \CaII\ NIR. Only in the SNe II does the \CaII\ NIR line appear unblended and resolved. As this line remains optically thick in the epochs considered, measurements made for other SN types should be considered largely uncertain. The outlying SN Ibn is SN 2006jc, whose spectra are noisy and the feature around 8600 \AA\ is particularly noisy and blended.}
    \label{fig:fwhms_ca}
\end{figure}

In the intensity-weighted plot, the FWHM values vary around the line of unity. The suggestion is that the emission of the \caiif\ and the `core' \Oneb\ occur in similar regions. In the flux-weighted plot, we can see that \caiif\ mostly originates at lower velocities than \Oneb.
The position of some SNe II are significantly affected based upon which metric is used because the broad and narrow components are of similar intensity, while in SE-SNe the broad component is always the strongest. Based on the size of the uncertainties calculated from our repeating of the fitting with different continuum positions, it is likely that some intrinsic property of the \Oneb\ distribution, such as multiple emitting regions, is sensitive to the chosen method of estimating the FWHM. 
For a couple of nearby and well-observed objects (e.g., SNe 1987A and 2004et), a single component fit replicates the \fwhmo\ results of the two-component fit. In these cases the average \fwhmo\ is $\sim 4000$ \kms.
For narrow-lined objects, a single component is insufficient to replicate the line shape due to the presence of excess flux, which may be continuum or evidence of a broad component. 
In these cases, however, the narrow component is clear and \fwhmo\ can even be estimated from measurements of this component alone.
Beyond this, there are a few objects where the line is not clear enough to fully discriminate between a narrow and broad component but is not well fit by a single component.

The right-hand panel of Fig.~\ref{fig:fwhms} shows that the \caiif\ 7291 \AA\ and \Oneb\ 6300 \AA\ flux-weighted time-averaged FWHM 
of the various SN types take a range of values, with the SNe II being the most separated in the FWHM parameter space. Another way of showing this is the density distributions shown in Fig.~\ref{fig:fwhm_dist}, where the stacked distributions from the right-hand panel of Fig.~\ref{fig:fwhms} are shown for the \CaII\ and \Oneb\ lines. These allow the median and intrinsic width of the SN classes to be studied with the statistics of each distribution given in Table~\ref{tab:fwhm_stats}.\footnote{An approximation of a Gaussian distribution has been used to estimate the widths (1-sigma uncertainties) that is generally reasonable based on the distributions in Fig.~\ref{fig:fwhm_dist} but not necessarily appropriate in all cases.}   For almost all transients, the \Oneb\ emitting region is more extended than that of \caiif\ (see further discussion in Section \ref{sec:emitting_region}). The distribution of FWHM$_{7921}$ for all but the SNe II is peaked in the range 4000 -- 6000 \kms. The SNe II are at lower velocities, with an average of $\sim2700$ \kms. For FWHM$_{6300}$, the SE-SNe are more tightly constrained than the SNe II. The order of increasing \Oneb\ 6300 \AA\ width for the IIb, Ib, Ic and Ic-BL is in agreement with the SE-SN results of \cite{2022arXiv220111467F}. We also find a similar trend for the \caiif\ 7291 \AA. The distributions of the SLSNe, Ca-rich events, and SNe Ibn, are severely affected by the weakness of the \Oneb\ line in some of the objects.

Figure~\ref{fig:fwhms_ca} shows the \CaII\ \fwhmf\ against \CaII\  \fwhmc, where the measurements are most consistent for the SNe II, suggesting that the emitting regions for the two \CaII\ lines in SN II are effectively the same within the uncertainties of the measurements. However, this does not necessarily imply that these lines are not forming in slightly different regions  \citep[cf.][]{Jerkstrand2017}, just that this method is too coarse to capture these nuances. For most SN types shown in Fig.~\ref{fig:fwhms_ca}, the lines are broad, shifted, and blended to varying degrees resulting in a departure from a one-to-one correspondence between the \CaII\ lines. The presence of [\CI] and [\FeII] in this region, combined with the scattering due to the large optical depth in the lines at these times that was discussed earlier, means the FWHM measurements of the \CaII\ NIR triplet are unreliable. All that is being measured is two broad features, one of which is poorly constrained, hence they fall into a cluster with no overall correlation.Therefore, in general, we find the \caiif\ values are more reliable than the \CaII\ NIR triplet values and give a better handle on the Ca distribution in the ejecta.

\begin{figure*}
    \centering
    \includegraphics{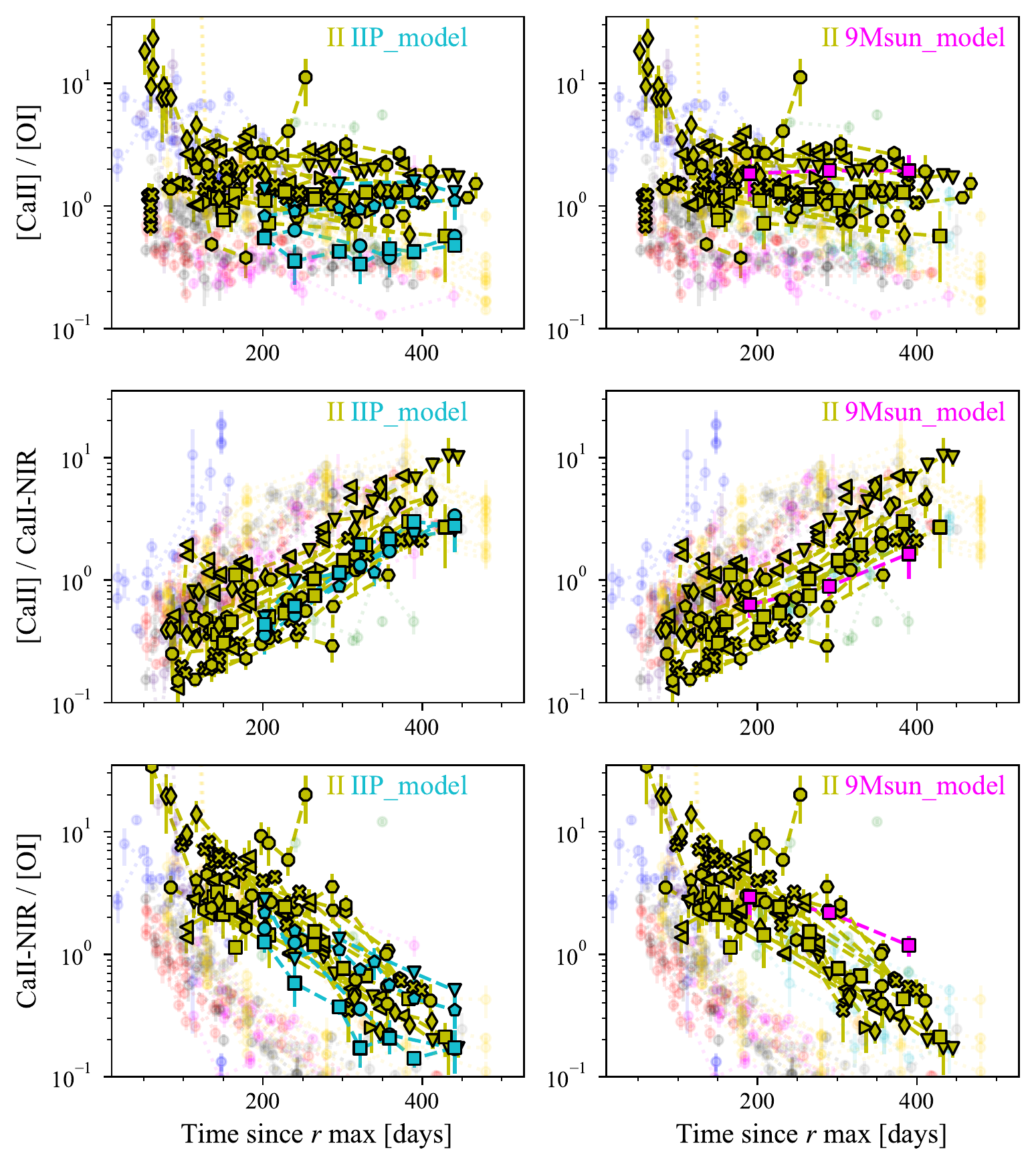}
    \caption{The observed line flux ratios of the SN II sample are shown in yellow with various shapes. They are compared with the 12 \msun\ (blue down triangles), 15 \msun\ (blue pentagons), 19 \msun\ (blue circles), and 25 \msun\ (blue squares) SN II nebular models of \citet{Jerkstrand2014} (left) and a 9 \msun\ model (pink squares) \citep[right,][]{Jerkstrand2018}. For comparison with the rest of the sample (non-SN II), the semi-transparent data points represent the SN types with different colours as described in the legend in Fig.~\ref{fig:caiif-O}.  The different input physics for the 9 \msun\ model compared to the higher mass ones is detailed in Section \ref{sec:iimodel}.} 
    \label{fig:IImodels}
\end{figure*}

\begin{figure*}
    \centering
    \includegraphics[scale=0.4]{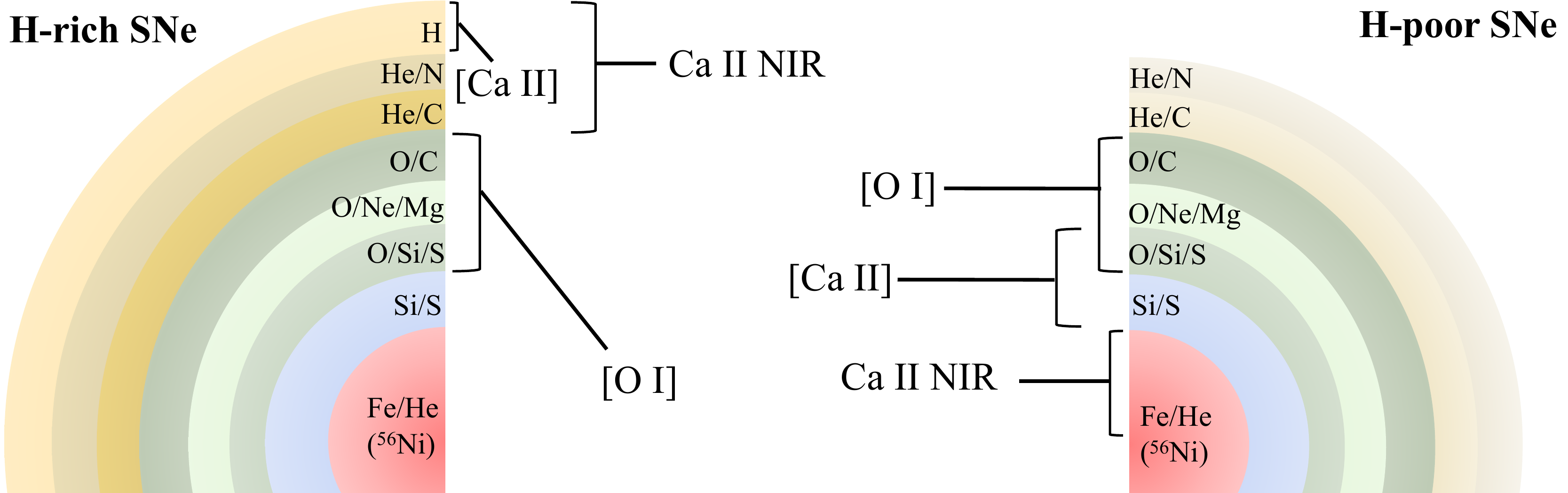}
    \caption{Cartoon representation of the canonical zonal structure for H-rich SN (left) and a H-poor SN (right) indicating where the \Oneb, \caiif, \CaII\ NIR line forming regions are located \citep{Li1993,Kozma1998,Dessart2011,Jerkstrand2012,Jerkstrand2015}. In a Type II SN the H envelope is intact, whereas for SE-SNe this is either almost entirely stripped (SN IIb) or is entirely stripped (SN Ib and Ic). The absence of the H envelope in SE-SNe means that \CaII\ emission comes from Ca synthesised in explosive burning \citep{Jerkstrand2015}, while Ca in the H envelope of a SN II is primordial \citep{Li1993}. The results of Section~\ref{sec:fwhm}, where \caiif\ \fwhmf\ $<$ \Oneb\ \fwhmo\ for SNe II, would suggest that the \caiif\ forms deeper in the ejecta compared with \Oneb\ in these objects, a situation as in the He-poor SNe. This is in some tension with the defined regions as represented here, and suggests the contribution from synthesised Ca may be significant also in Type II SNe \citep[see also][]{2021arXiv210513029D}.}
    \label{fig:cartoon}
\end{figure*}

\section{Discussion}\label{sec:discussion}

The measured properties, such as line-flux ratios and velocity widths, of late-time SN spectra can provide information on their underlying explosion and progenitor properties. \caiiflam\ was found to be the easiest line to fit in all cases, due to its strength and only mild contamination from other nearby sources of emission. 
Comparatively, \Oneb\ can be approximated by a two-component fit, one narrow and one broad, in $\sim$80 per cent of SNe studied.  However, it is possible to fit the \Oneb\ emission in some SNe II as a single Gaussian doublet, while in a small number (four events), a third contribution is required but is always weak and narrow. In all cases of multiple components, we find that the broad component dominates the total flux of the line. For the \CaII\ NIR emission, we find that the line can only be fit conclusively in SNe II. For other objects a mixture of line blending and the presence of other sources of emission in this region leads to large uncertainties on the contribution of \CaII\ NIR to the emission around 8600 \AA.

The line flux ratio curves of some SN types have been shown in the previous section to have distinct tracks as a function of time suggesting distinct progenitor characteristics. 
It has also been shown that the FWHM velocities of the lines show a trend between object types but not within object types, and that typically \Oneb\ emission is broader than \caiif, which implies an extended emission region in the ejecta. To further investigate the significance of these observational findings we compare our results to models of SNe II and IIb. We consider the accuracy of line-flux ratios in estimating core O masses and the \ek\ of the explosions. We also discuss the potential connections between SNe II and SE-SNe and between SNe Ic and SLSN-I.

\subsection{Comparing Type II SN data and models}\label{sec:iimodel}

To provide an accurate comparison between literature models of SNe II and our data, our line-fitting method is applied to model SN II spectra from \citet{Jerkstrand2014}, which are derived from the explosion models of \citet{Woosley2007}, and to the 9 \msun\ model from \citet{Jerkstrand2018}. The 9 \msun\ model differs from the higher mass models due to both the explosion physics employed (neutrino instead of piston driven), as well as assuming different mixing (no mixing instead of strong core mixing). It should, therefore, be considered as distinct from the other models.
The three line ratios (\fo, \fc, and \co) estimated from the models are plotted with the data for SNe II as a function of phase in Fig.~\ref{fig:IImodels}. 

The SN II nebular models presented in \cite{Jerkstrand2014} have masses of 12, 15, 19, and 25 \msun\ and are shown in the left-hand panels. The right-hand panels of Fig.~\ref{fig:IImodels} show the evolution of the 9 \msun\ model. The lower mass model follows a similar path to the higher mass ones and can be considered a lower limit on the expected line-flux ratio evolution. The models show a wide range of values in the \fo\ plane, and the size of the ratio is inversely proportional to the mass of the explosion model, with larger mass models having smaller ratios.
The tracks of the actual SNe are only located around the position of the 9, 12 and 15 \msun\ models or at larger ratios. No object follows the 19 and 25 \msun\ tracks.

In the \fc\ plane, the model flux ratios are bunched closely together.
As with \fo, the model and observed tracks only overlap for certain ratios. 
There is no correlation between the nearest \fo\ model track and the nearest \fc\ model track for the actual SNe. 

Finally, in the \co\ plane, the models again show a split based upon progenitor mass.
The time evolution is marginally slower for the models than for the observed data, unlike for \fo\ and \fc, but for the most part the actual and model tracks overlap. This reveals that \caiif\ emission is weaker in the models than in the observed spectra.

\subsection{Line profiles as a diagnostic of emission regions}
\label{sec:emitting_region}
Earlier studies \citep[e.g.,][]{Li1993,Kozma1998,Jerkstrand2012} demonstrated that the Ca which gives rise to \caiif\ emission in SNe II is primordial and found in the inner H envelope. However, \citet{Dessart2011} suggested that the \caiif\ emission around 300 d in SNe II comes primarily from the core, below the He layers. Their argument was based on the similarity of the \caiif\ emission line FWHM to that of \Oneb. 
They also suggested that emission from \CaII\ NIR is found all throughout the H envelope.
\citet{Dessart2020} investigated several different scenarios involving mixing of various elements into the H envelope of SN II progenitors.
They found that when \Nifs\ was mixed into the H envelope, it caused a Ca over-ionisation (to Ca III), which reduced the emission from \caiif. 
The results here show that the \fo\ ratio for SNe II in our sample are similar, with no significant outliers. This suggests that extreme mixing of \Nifs\ into the H envelope, to the extent that Ca II ionises to Ca III, does not appear to occur often.
Likewise, mixing of Ca from the Si-rich shell into the O shell during Si burning reduces emission from the \Oneb\ doublet because \caiif\ is the more effective coolant \citep{Dessart2018,Dessart2020}. This would lead to a much higher \fo\ ratio, which is not observed in the sample and suggests that extreme mixing does not occur.

In each plane
in Figs.~\ref{fig:caiif-O}, \ref{fig:caiif-canir}, and \ref{fig:canir-O}, the SNe II are separated from the SE-SNe to varying degrees. We consider now how much of this can be attributed to the presence of a H-envelope.  As SE-SNe lack a H envelope, the Ca responsible for \caiif\ emission is primarily expected to be found in the Si/S shell, a consequence of explosive O burning or incomplete Si burning, while Ca II NIR emission comes from Ca synthesised in the \Nifs\ shell \citep{Jerkstrand2015}.
Figure~\ref{fig:fwhm_dist} shows that the \caiif\ FWHM$_{7921}$ is, on average, lower than that of the \Oneb\ FWHM$_{6300}$ when using the weighted-flux value. This would suggest that the Ca responsible for this emission is coming from regions deeper than the O is.

Figure~\ref{fig:cartoon} demonstrates the canonical structure, and suggested line-forming regions of the \Oneb, \caiif, \CaII\ NIR features, for H-rich (SNe II) and H-poor (SE-SNe) events, as described in previous theoretical works \citep{Li1993,Kozma1998,Dessart2011,Jerkstrand2012,Jerkstrand2015}. The results presented here in terms of elemental distributions for H-poor SNe agree well with the previously presented theoretical structure. However, for the H-rich schema, our results based on an analysis of the FWHM of the lines suggest that the \caiif\ emission comes from regions deeper in the ejecta than the \Oneb\ emission.

\subsection{Comparison between Type IIb models and SE-SNe observations}
\begin{figure}
    \centering
    \includegraphics[scale=0.95]{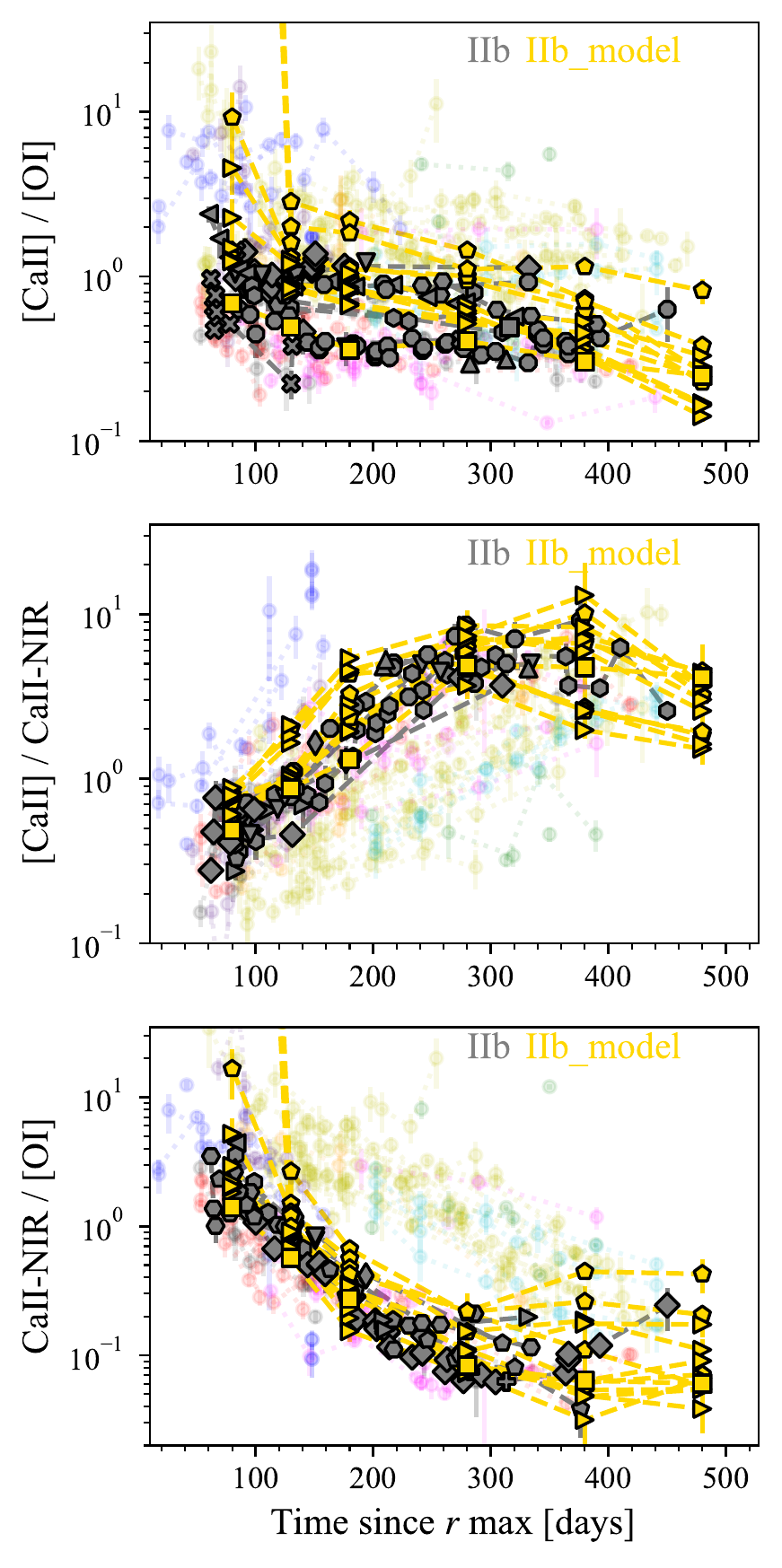}
    \caption{The SNe IIb ratios (dark grey) compared with the model nebular spectra (yellow) of \citet{Jerkstrand2015}. The nebular spectra are derived from explosion models of 12 \msun\ (pentagons), 13 \msun\ (right triangles), and 17 \msun\ (squares). These models also have varying parameters of mixing, energy deposition, and dust but it is the mass sequence that is the focus here. Unlike the SNe II, there is considerable overlap between the tracks of the models and the observed data. The semi-transparent points represent the SN sub-types with different colours as described in the legend in Fig.~\ref{fig:caiif-O}. } 
    \label{fig:IIbmodels}
\end{figure}

Figure~\ref{fig:IIbmodels} shows where the line ratios for the  \citet{Jerkstrand2015} models of SNe IIb sit in relation to the observed SNe IIb. 
These models are derived from the \citet{Woosley2007} explosion models, with envelopes artificially stripped, and feature stars with \mzams\ of 12, 13, and 17 \msun.
The 12 and 13 \msun\ models are further split based on a variety of parameters including mixing, clumping, dust, and molecular cooling \citep[see][for details]{Jerkstrand2015}. 
However, we are primarily concerned with the general tracks as a function of mass and so group the models by mass. The majority of studied events have evolving line-flux ratios that are well matched by the model paths.  SNe Ib and Ic are also shown in Fig.~\ref{fig:IIbmodels} as black and red semi-transparent points as in earlier figures (e.g.~see Fig.~\ref{fig:IImodels} for the full range covered by them) and are found to follow the general trend of the SN IIb models, especially the 17 \msun~model. However, there are SNe Ibc found at lower \fo\ than the SNe IIb.

\subsection{The relationship between SNe Ic and SLSN-I}
SLSNe-I are H/He-poor and are known for some spectroscopic similarity to SNe~Ic during their late photospheric phase \citep{2018ApJ...855....2Q}.
It has been demonstrated here that the SLSNe in the sample follow different tracks to those of the SE-SNe (see Fig.~\ref{fig:caiif-O}, \ref{fig:caiif-canir} and \ref{fig:canir-O}).
Two objects do eventually fall into the same position as the SNe Ibc in \fo\ but only at a later time.
One aspect to consider is the difference in time scale of the evolution. The typical half-width half-maximum rise time for He-poor SE-SNe is around 9 d \citep{Prentice2019}, for SLSNe this values is around 3--4 times longer \citep{DeCia2018}.
It is found that by scaling the given nebular phases of some SLSNe down by a factor of 3--4, three of the SLSNe (PTF12dam, \sn2015bn, \sn2016eay) are in the same position as the SE-SNe in each of the three ratios.
The exceptions are \sn2017egm and \sn2010md (marked in Fig.~\ref{fig:caiif-O}), which display extremely weak \Oneb\ emission and strong \caiif\ and \CaII\ NIR emission and indeed, the scaling would place them in a similar position to the Ca-rich transients (and the Type II, but these are in the very early nebular phases and not comparable with emission-line dominated spectra).
The \fc\ ratio for all SLSNe is $\sim 0.5$, considerably lower than for CC-SNe in general but in agreement with the ratio obtained for SLSNe in \citet[][]{2019ApJ...871..102N}. However, for PTF12dam, \sn2015bn, and \sn2016eay this is comparable to the SE-SNe if the time is scaled down by a factor of 3--4 and provides further evidence that these events are not incompatible with a common origin. Doing the same for \sn2017egm and \sn2010md places them with the Ca-rich transients, although this overlap may be purely coincidental.

\subsection{\fo\ as a proxy for core O mass and kinetic energy}
A key question we would like to address is if the \fo\ ratio is a reliable method of estimating the core O mass of the SNe in our sample. Increased O production scales strongly with increasing progenitor mass and therefore, the O mass is considered a good tracer of the progenitor mass \citep[e.g.][]{1995ApJS..101..181W}. The SNe II plotted in Fig.~\ref{fig:IImodels}  have \fo\ ratios that decrease for increasing progenitor mass and we find that the SNe II in our sample are most consistent with models with masses in the range of 9 to 15 \msun. The difference in the \co\ ratio between the models of different masses is smaller and suggests it is less suitable for linking to the progenitor masses, although the 25 \msun\ can be excluded as a good match for the objects in the sample based on this ratio alone. 

For the SNe IIb (Fig.~\ref{fig:IIbmodels}), there is more overlap in all three line ratios between the models of different masses (12, 13 and 17 \msun) due to the inclusion of additional physics (dust, mixing, clumping, molecular cooling). In general, we find that the model tracks overlap well with the data, and show very similar evolution with time. This suggests that SNe IIb can be explained by the progenitors of \citet{Jerkstrand2015} with masses of 12 - 17 \msun, consistent with their results but using the easier method of line-flux ratios that is less sensitive to spectral calibration uncertainties. 

However, when we compare the measured \fo\ ratios to the \mej\ calculated from detailed light-curve modelling (Fig.~\ref{fig:mass-eom}), we find no connection between \mej\ and the line ratios SE-SNe. One possibility is that another power source aside from the radioactive decay of \Nifs, probably located to the interior of the ejecta, is affecting the transient energetics or dynamics and in turn decoupling the relationship between light curve width and ejecta velocity that is normally used to calculate \mej\ from the light curve \citep[e.g.,][]{1982ApJ...253..785A}. A good example of this would be the presence a magnetar, which has been invoked in order to explain the long-lived and luminous light curves of SLSNe.
The application of a magnetar model instead of a \Nifs\ model greatly reduces the required ejecta mass \citep[e.g.,][]{2013ApJ...770..128I} because the shape of the magnetar luminosity curve is variable, unlike that for radioactive decay which is fixed.
Unusual SE-SN events, such as the double-peaked Type Ib \sn2005bf \citep{2005ApJ...633L..97T} and the long-lived and luminous Ic \sn2019cri \citep{2021MNRAS.508.4342P} demonstrate that some spectroscopically normal SE-SNe display light curves that cannot be powered solely by the \Nifs\ decay chain.
We also tested the \fo\ ratio against the specific kinetic energy \eom\ (\ek\ per unit mass in units of \eomu), see Fig.~\ref{fig:mass-eom}. No sequence was found within the SN subtypes, particularly within the SNe Ic, where significant differences in \eom\ are known to exist \citep{Prentice2017}.

\section{Conclusions}\label{sec:conclusions}
In this work, we have presented the results of a study investigating the simple approximation of the SN nebular emission lines \Oneblam, \caiiflam, and \CaII\ NIR as a series of Gaussian line profiles for a large sample of SNe displaying these lines. 
The peak intensity, flux, and FWHM were extracted for each studied lines.
We demonstrated that the various SN types follow differing tracks in the flux ratio curves of these emission lines.
For \fo, we find that SNe II typically have the largest ratio of the CC-SNe, followed by SNe IIb and then SNe Ibc.
This is partially related to the fact that a significant fraction of the \caiif\ emission in H-rich SNe comes from primordial Ca in the H envelope. Without this envelope, as in the case of SNe Ib and Ic, \caiif\ emission arises from the ashes of explosive burning.

The \fo\ curves of SE-SNe were compared with the physical properties \mej\ and \eom\ (obtained from literature modelling of their light curves) to check for any correlations in ratio value and physical values, none was found.
This means that the \fo\ curves of highly energetic objects and higher ejecta mass objects were found to occupy the same parameter space as low-energy and low-mass objects. As O mass scales with progenitor mass, this result suggests that the progenitors of these SNe are similar and that their differences in light curve shape, or energetics, could relate to the powering mechanism, which in turn breaks the coupling between light curve width and ejecta mass found in analytical light curve models \citep[e.g.,][]{1982ApJ...253..785A}. As a consequence, SE-SNe may not have as high ejecta masses as suggested by their light curve modelling.

The results of the estimated line parameters were compared to model nebular spectra of SNe II and SNe IIb and a preference was found for SNe II to follow the tracks of models with \mzams\ $<15$ \msun, and none with \mzams\
 $>19$ \msun\ models. SNe IIb favour models with \mzams\ $<17$ \msun, although a large range of parameters used to derive the models results in overlap between the model ratio curves.

Comparison of the average FWHM measurement of \Oneb\ and \caiif\ shows that the latter are typically narrower than the former, which suggests that typical line-forming region of \caiif\ is found deeper in the ejecta that \Oneb, which is contrary to the standard models (cf. Fig.~\ref{fig:cartoon}). 
Additionally, we have been able to show that in SN II, the FWHM of \caiif\ and \CaII\ NIR are approximately equal, which again suggests similar locations to the line-forming regions.
The results here also show that the flux ratio curves of \caiif\ and \Oneb\ of different SN types are found in well-defined regions. There is no indication of extreme suppression of \caiif\ in SNe II that would be expected if there was significant mixing of \Nifs\ into the H envelope. 
There is also no clear indication of an enhancement of \caiif\ at the expense of \Oneb\ as would be the case if Ca was mixed in the O shell in the progenitor.

Finally, we have found that the line-flux ratios of SLSNe-I and SE-SNe are in agreement when the relative phases with respect to maximum light are scaled down based on their relative rise times. This provides evidence, in agreement with suggestions in the literature, that there is a connection between the origins of SLSNe-I and SE-SNe. However, this is not the case for all SLSNe-I events studied. 

\section*{Acknowledgements}
KM and SJP are supported by H2020 European Research Council (ERC) Starting Grant no.~758638 (SUPERSTARS).
AJ acknowledges funding from ERC Starting Grant no.~803189 (SUPERSPEC) and the Swedish National Research Council (Etableringsbidrag 2018-03799).

\section*{Data Availability}
The data used is publicly available through WISeREP.



\bibliographystyle{mnras}
\bibliography{nbib} 



\appendix
\section{Table of average FWHM}
\begin{table*}
    \centering
    \caption{$100-200$ day time-averaged FWHM information for each object. The peak measurement FWHM$_\textrm{O}$ (peak) is the FWHM of the \Oneb\ component with the greatest peak intensity, while FWHM$_\textrm{O}$ (flux) is measured from the \Oneb\ component with the largest flux. The flux weighted value, FWHM$_\textrm{O}$ (flux weighted), refers to the \Oneb\ component with the velocity obtained by combining the \Oneb\ fluxes weighted in proportion to their contribution to the total flux. The \caiif\ only has one component so a single value is included (FWHM$_\textrm{Ca}$).   The classification of SN~Ic is the scheme described in the caption of Table \ref{tab:sample}.}  
    \begin{tabular}{llllll}
SN  &  Type  &  FWHM$_\textrm{O}$ (peak)  &  FWHM$_\textrm{O}$ (flux)  &  FWHM$_\textrm{O}$ (flux weighted)  & FWHM$_\textrm{Ca}$\\
  &   & (\kms) &   (\kms)   &   (\kms)   &  (\kms) \\
\hline
1987A & II & $3528\pm^{36}_{43}$ & $3528\pm^{36}_{37}$ & $3125\pm^{26}_{51}$ & $2648\pm^{36}_{99}$\\
1990B & Ic & $1104\pm^{706}_{706}$ & $8015\pm^{1749}_{1834}$ & $6431\pm^{1254}_{1186}$ & $4301\pm^{414}_{414}$\\
1990E & II & $1076\pm^{27}_{95}$ & $4489\pm^{274}_{187}$ & $3614\pm^{67}_{140}$ & $3623\pm^{336}_{480}$\\
1990K & II & $6121\pm^{256}_{112}$ & $6121\pm^{256}_{112}$ & $6121\pm^{256}_{112}$ & $4849\pm^{403}_{445}$\\
1992H & II & $5388\pm^{571}_{668}$ & $5904\pm^{552}_{625}$ & $5334\pm^{107}_{304}$ & $3410\pm^{105}_{198}$\\
1993J & IIb & $2100\pm^{307}_{89}$ & $10258\pm^{189}_{305}$ & $5850\pm^{208}_{248}$ & $3841\pm^{444}_{46}$\\
1993K & II & $6468\pm^{184}_{233}$ & $6468\pm^{184}_{193}$ & $5624\pm^{96}_{113}$ & $2768\pm^{27}_{30}$\\
1994I & Ic & $6823\pm^{74}_{85}$ & $6823\pm^{74}_{85}$ & $6665\pm^{31}_{82}$ & $5795\pm^{123}_{218}$\\
1996cb & IIb & $5394\pm^{305}_{2441}$ & $6696\pm^{230}_{638}$ & $5865\pm^{386}_{208}$ & $4359\pm^{188}_{54}$\\
1997ef & Ic-BL & $6995\pm^{80}_{299}$ & $6995\pm^{80}_{299}$ & $6737\pm^{98}_{255}$ & $5224\pm^{153}_{76}$\\
1998bw & Ic-BL & $7164\pm^{27}_{35}$ & $7164\pm^{27}_{35}$ & $6731\pm^{123}_{77}$ & $5355\pm^{158}_{248}$\\
1999em & II & $4896\pm^{954}_{983}$ & $4896\pm^{954}_{983}$ & $4565\pm^{949}_{979}$ & $2643\pm^{129}_{445}$\\
2001ig & IIb & $2699\pm^{2526}_{2526}$ & $5292\pm^{279}_{293}$ & $5168\pm^{296}_{280}$ & $4571\pm^{591}_{591}$\\
2002ap & Ic-BL & $8895\pm^{868}_{1196}$ & $9813\pm^{41}_{921}$ & $8450\pm^{201}_{564}$ & $5608\pm^{670}_{406}$\\
2002hh & II & $3557\pm^{181}_{355}$ & $3557\pm^{181}_{355}$ & - & $3052\pm^{367}_{19}$\\
2003B & II & $1595\pm^{556}_{556}$ & $6874\pm^{653}_{685}$ & $4567\pm^{693}_{655}$ & $2099\pm^{296}_{296}$\\
2003bg & IIb & $4831\pm^{27}_{1096}$ & $4873\pm^{240}_{28}$ & $4114\pm^{89}_{63}$ & $4812\pm^{325}_{239}$\\
2003hn & II & $4111\pm^{3327}_{3327}$ & $6736\pm^{824}_{864}$ & $5769\pm^{854}_{808}$ & $3018\pm^{96}_{101}$\\
2004aw & Ic & $6353\pm^{168}_{546}$ & $6353\pm^{168}_{546}$ & $5562\pm^{406}_{382}$ & $3565\pm^{790}_{400}$\\
2004dj & II & $3423\pm^{83}_{92}$ & $3423\pm^{83}_{22}$ & $3147\pm^{104}_{129}$ & $2768\pm^{148}_{329}$\\
2004et & II & $3753\pm^{286}_{124}$ & $3793\pm^{246}_{117}$ & $3677\pm^{219}_{172}$ & $2888\pm^{45}_{223}$\\
2004gk & Ic & $7030\pm^{45}_{246}$ & $7030\pm^{45}_{246}$ & $6879\pm^{42}_{306}$ & $4784\pm^{244}_{172}$\\
2004gq & Ib & $7490\pm^{470}_{470}$ & $7490\pm^{470}_{470}$ & $7279\pm^{586}_{554}$ & $4767\pm^{76}_{179}$\\
2004gt & Ib & $5623\pm^{345}_{345}$ & $5623\pm^{345}_{345}$ & $5349\pm^{429}_{406}$ & $4471\pm^{519}_{519}$\\
2005E & Ca-rich & $6684\pm^{2941}_{2941}$ & $6684\pm^{2941}_{2941}$ & $6160\pm^{3459}_{3270}$ & $3710\pm^{484}_{484}$\\
2005ay & II & $1204\pm^{525}_{525}$ & $5347\pm^{616}_{646}$ & $4250\pm^{654}_{618}$ & $1884\pm^{593}_{593}$\\
2005bf & Ib & $6205\pm^{435}_{435}$ & $6205\pm^{435}_{435}$ & $5666\pm^{468}_{442}$ & $4588\pm^{312}_{312}$\\
2005cs & II & $1885\pm^{81}_{88}$ & $1885\pm^{81}_{88}$ & $1885\pm^{81}_{88}$ & $1645\pm^{41}_{149}$\\
2006aj & Ic-BL & $1401\pm^{2186}_{19}$ & $8232\pm^{671}_{250}$ & $7647\pm^{421}_{386}$ & $4088\pm^{522}_{1061}$\\
2006bp & II & $5265\pm^{663}_{663}$ & $5265\pm^{663}_{663}$ & $4815\pm^{826}_{781}$ & $2805\pm^{513}_{513}$\\
2006el & IIb & $7231\pm^{405}_{405}$ & $7231\pm^{405}_{405}$ & $6475\pm^{504}_{477}$ & $4635\pm^{422}_{422}$\\
2006jc & Ibn & $1065\pm^{51}_{78}$ & $7662\pm^{491}_{1083}$ & $6357\pm^{496}_{841}$ & $1928\pm^{465}_{242}$\\
2007I & Ic-BL & $6388\pm^{468}_{468}$ & $6388\pm^{468}_{468}$ & $5888\pm^{582}_{551}$ & $5537\pm^{376}_{376}$\\
2007aa & II & $887\pm^{288}_{288}$ & $5783\pm^{338}_{355}$ & $4615\pm^{359}_{340}$ & $1761\pm^{290}_{290}$\\
2007gr & Ic & $6939\pm^{45}_{150}$ & $6980\pm^{58}_{50}$ & $6796\pm^{46}_{86}$ & $5032\pm^{131}_{98}$\\
2007ke & Ca-rich & $2303\pm^{445}_{445}$ & $7094\pm^{522}_{548}$ & $4938\pm^{554}_{524}$ & $3826\pm^{251}_{251}$\\
2008D & Ib & $4458\pm^{2229}_{2229}$ & $6513\pm^{311}_{326}$ & $5518\pm^{397}_{375}$ & $5502\pm^{717}_{717}$\\
2008ax & IIb & $5957\pm^{674}_{1321}$ & $6669\pm^{189}_{294}$ & $5633\pm^{494}_{570}$ & $4056\pm^{102}_{242}$\\
2008bk & II & $1004\pm^{683}_{114}$ & $3177\pm^{1283}_{713}$ & $3048\pm^{426}_{287}$ & $1243\pm^{26}_{75}$\\
2008bo & IIb & $2213\pm^{397}_{414}$ & $6603\pm^{3785}_{3943}$ & $4524\pm^{837}_{883}$ & $3882\pm^{332}_{349}$\\
2009N & II & $1068\pm^{29}_{24}$ & $6307\pm^{464}_{1341}$ & $4719\pm^{747}_{1008}$ & $1093\pm^{47}_{53}$\\
2009ib & II & $737\pm^{79}_{82}$ & $5638\pm^{441}_{505}$ & $4635\pm^{659}_{688}$ & $2321\pm^{56}_{58}$\\
2009jf & Ib & $6095\pm^{892}_{859}$ & $6950\pm^{151}_{192}$ & $5972\pm^{587}_{557}$ & $4165\pm^{533}_{581}$\\
2010et & Ca-rich & $7067\pm^{35}_{337}$ & $7067\pm^{35}_{337}$ & $6826\pm^{90}_{528}$ & $4755\pm^{206}_{325}$\\
2010lp & Ia & $2973\pm^{583}_{583}$ & $2973\pm^{583}_{583}$ & $2829\pm^{726}_{686}$ & $3726\pm^{447}_{447}$\\
2010md & SLSN* & $4216\pm^{362}_{362}$ & $4216\pm^{362}_{362}$ & $4216\pm^{362}_{362}$ & $4802\pm^{344}_{344}$\\
2011bm & Ic & $6697\pm^{259}_{126}$ & $6697\pm^{259}_{126}$ & $6418\pm^{319}_{43}$ & $7027\pm^{90}_{66}$\\
2011dh & IIb & $5628\pm^{277}_{521}$ & $5760\pm^{731}_{156}$ & $5285\pm^{227}_{470}$ & $3286\pm^{563}_{230}$\\
2011ei & IIb & $5920\pm^{169}_{405}$ & $5920\pm^{169}_{405}$ & $5720\pm^{25}_{367}$ & $3961\pm^{398}_{47}$\\
2011fu & IIb & $6775\pm^{213}_{659}$ & $6775\pm^{213}_{659}$ & $6756\pm^{112}_{672}$ & $4723\pm^{101}_{683}$\\
2011hs & IIb & $3812\pm^{19}_{29}$ & $3812\pm^{19}_{29}$ & $3497\pm^{2}_{12}$ & $3327\pm^{35}_{39}$\\
2012ap & Ic-BL & $7872\pm^{102}_{207}$ & $8031\pm^{267}_{108}$ & $7251\pm^{104}_{251}$ & $4200\pm^{177}_{77}$\\
2012aw & II & $3443\pm^{96}_{393}$ & $3443\pm^{96}_{33}$ & $3287\pm^{46}_{307}$ & $2366\pm^{153}_{55}$\\
2012ec & II & $2419\pm^{1183}_{1183}$ & $3462\pm^{165}_{173}$ & $3000\pm^{415}_{392}$ & $2752\pm^{544}_{544}$\\
2012hn & Ca-rich & $1988\pm^{416}_{416}$ & $2655\pm^{488}_{512}$ & $2348\pm^{518}_{490}$ & $5523\pm^{361}_{361}$\\

\hline
    \end{tabular}
    \label{tab:my_label}
\end{table*}

\begin{table*}
    \centering
    \contcaption{$100-200$ day time-averaged FWHM values.}
    \begin{tabular}{llllll}

SN  &  Type  &  FWHM$_\textrm{O}$ (peak)  &  FWHM$_\textrm{O}$ (flux)  &  FWHM$_\textrm{O}$ (flux weighted)  & FWHM$_\textrm{Ca}$\\
  &   & (\kms) &   (\kms)   &   (\kms)   &  (\kms) \\
\hline
2013ab & II & $1949\pm^{106}_{119}$ & $5589\pm^{187}_{202}$ & $4255\pm^{92}_{96}$ & $2761\pm^{166}_{115}$\\
2013ak & IIb & $7412\pm^{246}_{246}$ & $7412\pm^{246}_{246}$ & $6658\pm^{307}_{290}$ & $2161\pm^{668}_{668}$\\
2013am & II & $768\pm^{138}_{19}$ & $5115\pm^{181}_{1496}$ & $3908\pm^{56}_{271}$ & $939\pm^{134}_{17}$\\
2013bb & IIb & $5710\pm^{420}_{420}$ & $5710\pm^{420}_{420}$ & $5230\pm^{524}_{495}$ & $4395\pm^{441}_{441}$\\
2013ej & II & $3516\pm^{90}_{232}$ & $3572\pm^{56}_{62}$ & $3176\pm^{189}_{211}$ & $4081\pm^{1270}_{187}$\\
2013ge & Ic & $6802\pm^{171}_{356}$ & $6802\pm^{171}_{356}$ & $6371\pm^{438}_{185}$ & $4139\pm^{255}_{39}$\\
2014G & II & $1921\pm^{233}_{153}$ & $4852\pm^{1728}_{2263}$ & $4726\pm^{503}_{669}$ & $3807\pm^{449}_{566}$\\
2014L & Ic & $6080\pm^{433}_{49}$ & $6080\pm^{433}_{49}$ & $5863\pm^{241}_{63}$ & $4227\pm^{66}_{66}$\\
2015G & Ibn & $6287\pm^{209}_{263}$ & $6545\pm^{38}_{44}$ & $5783\pm^{74}_{87}$ & $3260\pm^{276}_{299}$\\
2015ah & Ib & $1440\pm^{776}_{776}$ & $7650\pm^{911}_{955}$ & $6049\pm^{966}_{914}$ & $4642\pm^{688}_{688}$\\
2015ap & Ib & $7113\pm^{89}_{52}$ & $7113\pm^{89}_{52}$ & $6738\pm^{69}_{68}$ & $5050\pm^{357}_{400}$\\
2015bn & SLSN-I & $7783\pm^{266}_{137}$ & $8194\pm^{522}_{397}$ & $7843\pm^{403}_{168}$ & $5779\pm^{317}_{279}$\\
2016coi & Ic-BL & $3337\pm^{68}_{288}$ & $5999\pm^{4275}_{473}$ & $5995\pm^{228}_{473}$ & $5856\pm^{184}_{394}$\\
2016eay & SLSN-I & $8376\pm^{312}_{395}$ & $8868\pm^{161}_{174}$ & $8334\pm^{182}_{194}$ & $3741\pm^{196}_{58}$\\
2016hgs & Ca-rich & $3501\pm^{1196}_{1196}$ & $3501\pm^{1196}_{1196}$ & $3339\pm^{1288}_{1218}$ & $5808\pm^{688}_{688}$\\
2016iae & Ic & $7149\pm^{95}_{95}$ & $7149\pm^{95}_{95}$ & $6980\pm^{148}_{140}$ & $5071\pm^{974}_{974}$\\
2016jdw & Ib & $5424\pm^{465}_{465}$ & $5424\pm^{465}_{465}$ & $5414\pm^{579}_{548}$ & $5097\pm^{1874}_{1874}$\\
2017bgu & Ib & $5951\pm^{1023}_{1023}$ & $5951\pm^{1023}_{1023}$ & $5806\pm^{1173}_{1109}$ & $4768\pm^{751}_{751}$\\
2017egm & SLSN-I & $6228\pm^{308}_{308}$ & $6228\pm^{308}_{308}$ & $5853\pm^{384}_{363}$ & $3830\pm^{313}_{313}$\\
2017ein & Ic & $6714\pm^{787}_{787}$ & $6714\pm^{787}_{787}$ & $6239\pm^{906}_{857}$ & $6126\pm^{918}_{918}$\\
2018gjx & Ibn & $2317\pm^{251}_{355}$ & $2317\pm^{251}_{355}$ & - & $4027\pm^{286}_{526}$\\
2019yz & Ic & $7108\pm^{70}_{70}$ & $7108\pm^{70}_{70}$ & $6869\pm^{112}_{106}$ & $4386\pm^{306}_{306}$\\
OGLE-2012-SN-006 & Ibn & $7808\pm^{558}_{515}$ & $7808\pm^{558}_{515}$ & $6978\pm^{672}_{66}$ & $4456\pm^{349}_{868}$\\
PS15bgt & IIb & $5488\pm^{183}_{3267}$ & $5970\pm^{403}_{133}$ & $5524\pm^{224}_{16}$ & $4699\pm^{58}_{598}$\\
PTF09dav & Ca-rich & $11801\pm^{1531}_{1531}$ & $11801\pm^{1531}_{1531}$ & $10978\pm^{1906}_{1802}$ & $5510\pm^{304}_{304}$\\
PTF11bij & Ca-rich & $2589\pm^{732}_{732}$ & $2589\pm^{732}_{732}$ & $2286\pm^{911}_{862}$ & $3711\pm^{351}_{351}$\\
PTF11kmb & Ca-rich & $4560\pm^{20}_{25}$ & $4560\pm^{20}_{25}$ & $4213\pm^{25}_{32}$ & $3951\pm^{7}_{11}$\\
PTF12bho & Ca-rich & $8204\pm^{612}_{612}$ & $8204\pm^{612}_{612}$ & $7551\pm^{763}_{721}$ & $5601\pm^{691}_{691}$\\
PTF12dam & SLSN-I & $4111\pm^{3699}_{3699}$ & $9073\pm^{1481}_{1554}$ & $8791\pm^{1311}_{1240}$ & $4354\pm^{260}_{260}$\\
iPTF13bvn & Ib & $2289\pm^{262}_{262}$ & $2289\pm^{262}_{262}$ & $2396\pm^{327}_{309}$ & $3605\pm^{977}_{977}$\\
iPTF15eqv & Ca-rich & $6528\pm^{120}_{4893}$ & $7925\pm^{415}_{352}$ & $7374\pm^{178}_{938}$ & $4089\pm^{946}_{53}$\\

\hline
\multicolumn{6}{p{\textwidth}}{* \sn2010md is an example of a SLSN-IIb  \citep[see e.g.][]{2018ApJ...855....2Q} .} 
     \end{tabular}
    \label{tab:FWHM measurements}
\end{table*}


\bsp	
\label{lastpage}
\end{document}